\documentclass[aps,prb,reprint,superscriptaddress]{revtex4-1}
\usepackage[dvipsnames]{xcolor}
\usepackage{graphicx}
\usepackage{amsmath,amssymb}
\usepackage{bm}
\usepackage{braket}
\usepackage{lipsum}
\usepackage{makerobust}
\usepackage{simpler-wick}
\usepackage{comment}
\usepackage[noabbrev]{cleveref}

\newcommand*\circled[1]{\tikz[baseline=(char.base)]{
    \node[shape=circle,draw,inner sep=1pt] (char) {#1};}}
\MakeRobustCommand\circled

\newcommand{\makeauthor}[2]{\newcommand{#1}[1]{{%
  \sffamily\color{#2}{%
    \bfseries\begingroup\escapechar=-1\edef\x{\endgroup\string#1}\x:%
  } ##1}}%
  \MakeRobustCommand#1}
\makeauthor{\eric}{Plum}
\makeauthor{\dc}{magenta}
\makeauthor{\sr}{blue}
\makeauthor{\Fig}{red}

\begin{document}

\renewcommand{\vec}[1]{\bm{#1}}
\newcommand{\up}{{\uparrow}}
\newcommand{\dw}{{\downarrow}}
\newcommand{\pa}{{\partial}}
\newcommand{\pd}{{\phantom{\dagger}}}
\newcommand{\bs}[1]{\boldsymbol{#1}}
\newcommand{\add}[1]{{{\color{blue}#1}}}
\newcommand{\todo}[1]{{\textbf{\color{red}ToDo: #1}}}
\newcommand{\tbr}[1]{{\textbf{\color{red}\underline{ToBeRemoved:} #1}}}
\newcommand{\eps}{{\varepsilon}}
\newcommand{\nn}{\nonumber}
\def\ie{\emph{i.e.},\ }
\def\eg{\emph{e.g.},\ }
\def\ea{\emph{et. al.}\ }
\def\cf{\emph{cf.}\ }

\newcommand{\brap}[1]{{\bra{#1}_{\rm phys}}}
\newcommand{\bral}[1]{{\bra{#1}_{\rm log}}}
\newcommand{\ketp}[1]{{\ket{#1}_{\rm phys}}}
\newcommand{\ketl}[1]{{\ket{#1}_{\rm log}}}
\newcommand{\braketp}[1]{{\braket{#1}_{\rm phys}}}
\newcommand{\braketl}[1]{{\braket{#1}_{\rm log}}}

\graphicspath{{./}{./figures/}}



\title{Interplay of magnetic textures with spin-orbit coupled substrates}

\author{Zachary Llewellyn}
\affiliation{School of Physics, University of Melbourne, Parkville, VIC 3010, Australia}
\author{Eric Mascot}
\affiliation{School of Physics, University of Melbourne, Parkville, VIC 3010, Australia}
\author{Oleg A. Tretiakov}
\affiliation{School of Physics, University of New South Wales, Sydney, NSW 2033, Australia}
\author{Stephan Rachel}
\affiliation{School of Physics, University of Melbourne, Parkville, VIC 3010, Australia}
\noaffiliation

\date{\today}

\begin{abstract}
Magnetic textures such as skyrmions in thin films grown on substrates possess significant technological potential. Inhomogeneous magnetic structures can be described as homogeneous ferromagnetic order in the presence of anisotropic spin-orbit coupling (SOC). It remains unexplored, however, how this {\it induced} SOC stemming from the magnetic textures interacts with the SOC of the substrate. Here we show that these two contributions to SOC are in general {\it not} additive. We demonstrate this by employing a spintronics gauge theory. We further compute local currents which, when considered in the proper frame, match the spintronics gauge theory results. Finally, we analyze global transport quantities and show that they substantiate our previous results quantitatively. The implications for skyrmionics as well as topological superconductivity are discussed.
\end{abstract}

\maketitle


%
%

\section{Introduction \label{Section I}}

    In recent times, magnetic skyrmions have emerged as a key field of research in spintronics\,\cite{RevModPhys.76.323, HIROHATA2020166711, WANG2022169905, 10.1063/5.0072735, Song_2020} and topological superconductivity\,\cite{Mascot2021-hz, brüning2024noncollinearpathtopologicalsuperconductivity, petrović2024colloquiumquantumpropertiesfunctionalities, PhysRevB.102.224501, PhysRevB.100.064504, PhysRevB.105.224509, Lo_Conte2024-bs} with many applications for spin-based technologies.
    Due to their topological protection \cite{Je2020-ze, PhysRevX.13.041027}, enhanced thermal stability \cite{PhysRevB.88.195137, PhysRevB.96.134420}, and small physical size, magnetic skyrmions offer unique opportunities to create new spintronic technologies \cite{Gobel2021}. Furthermore, through coupling to superconducting materials, there are opportunities for realizing exotic superconducting phases in skyrmion crystal structures\,\cite{PhysRevB.93.224505, PhysRevB.100.064504, Mascot2021-hz}.
    One of these spintronic technologies is the use of skyrmions for data storage purposes, which have made significant experimental progress towards realization \cite{Wiesendanger2016-gn, ZHAO20242370}, through the encoding of information into topologically distinct skyrmion states.

    In magnetic materials with broken inversion symmetry, atoms with strong spin-orbit coupling (SOC) can mediate an antisymmetric exchange interaction between spins called the Dzyaloshinskii-Moriya interaction (DMI). The DMI favors the non-collinear arrangement of neighboring spins. The competing behavior between the DMI and exchange coupling, which favors collinear alignment, leads to new, exotic magnetic textures from spin spirals to skyrmions.
    For slowly varying magnetic spin textures, such as skyrmions, it is well known that itinerant electrons' spin tends to align with the magnetic texture. This process leads to the formation of induced electromagnetic fields that act on the itinerant electrons\,\cite{GEVolovik_1987, PhysRevB.57.R3213, PhysRevLett.98.246601, PhysRevLett.107.136804, Schulz2012-bz}. Through careful algebraic manipulation\,\cite{PhysRevB.102.224501} it is possible to represent these induced fields as induced adiabatic magnetic fields, SOC, and local chemical potentials.
    These induced electromagnetic fields have been previously explored theoretically in terms of Hall and spin Hall responses of skyrmionic systems\,\cite{PhysRevB.99.184427, PhysRevApplied.12.054032, PhysRevB.95.064426, PhysRevB.92.024411}.

    It is known that intrinsic SOC induces additional electromagnetic fields, which manifest as a spin-motive force\,\cite{TAN20201} leading to spin currents\,\cite{PhysRevLett.95.187203,TAN20201}.
    These SOC fields have previously been paired with spin texture fields in order to explore new phenomenon such as the suppression of the skyrmion Hall effect in magnetic textures coupled to SOC materials\,\cite{PhysRevApplied.12.054032}. Similarly, there have been theoretical attempts to mix nontrivial spin textures and intrinsic SOC for the purpose of hosting Majorana zero modes (MZMs) in both spin spiral\,\cite{PhysRevB.102.224501} and skyrmion\,\cite{PhysRevB.93.224505, PhysRevB.100.064504} magnet-superconductor hybrid systems.
    Along with this, it is possible to stabilize Bloch-type skyrmions on chiral magnets with Dresselhaus SOC\,\cite{doi:10.1126/science.1166767, Yu2010-ca, doi:10.1126/science.1214143} and N\'eel-type skyrmions on polar magnets with Rashba SOC\,\cite{Kezsmarki2015-mv, PhysRevLett.119.237201} further hinting at interesting behavior between skyrmion spin textures and substrate Rashba and Dresselhaus SOC.
    However, what is largely missing in the literature is an understanding of the domain-agnostic effects of SOC-skyrmion interactions on both the local and mesoscopic scale.

    In this paper, we investigate the interaction of substrate SOC and skyrmionic spin textures.
    Hereby we focus on both the atomic bond level and the mesoscopic scattering level. Specifically, we investigate how the interaction causes nontrivial induced SOC and adiabatic magnetic fields along with the symmetries that exist between the Hall responses of different interactions of the intrinsic and induced SOCs.
    Through the use of spintronic gauge theory we provide an analytical form for the induced phenomenon occurring from the interaction. We then corroborate these results through the use of quantum transport measurements to develop a new induced SOC marker to directly observe the interaction. From these bond-level results we find that the induced SOC has a non-additive relationship compared to if you considered the induced SOC from the skyrmion and the substrate SOC separately.
    Finally, we consider a four-terminal system hosting a magnetic skyrmion to investigate the mesoscopic scattering properties of a magnetic skyrmion coupled to a SOC substrate, from which we find pairings between different SOC-skyrmion systems.

    The remainder of the paper is structured as follows: In Section\,\ref{Section II} we investigate the interaction of SOC and skyrmions using spintronic gauge theory. A new marker, a local bond current, for induced SOC is introduced and then compared with the results of the previous Section in Section\,\ref{Section III}. The mesoscopic scattering effects of the interaction are analyzed using the Landauer-B\"uttiker formalism in Section\,\ref{Section IV}. In Section\,\ref{Section V} we provide a conclusion and an outlook.
    

%
%
\section{Spintronic Gauge Theory \label{Section II}}


\subsection{Continuum Theory \label{Section IIA}}

We consider a two-dimensional electron gas (2DEG) coupled with a magnetic material in the presence of intrinsic Rashba and Dresselhaus SOC. The general single-particle Hamiltonian for the continuum system has the form
\begin{equation}
    \mathcal{H} = \frac{\textbf{p}^{2}}{2 m^{*}} - \mu - \mathcal{H}_{\text{R}} - \mathcal{H}_{\text{D}} + J \textbf{m}\cdot\boldsymbol{\sigma},
    \label{continuum-hamiltonian}
\end{equation}
where the Rashba and Dresselhaus SOC terms, $\mathcal{H}_{R}$ and $\mathcal{H}_{D}$, respectively, are given by
\begin{equation}
    \begin{split}
        \mathcal{H}_{\text{R}} &= \frac{\alpha_{0}}{\hbar}(p_{x}\sigma_{y} - p_{y}\sigma_{x}),\\[5pt]
        \mathcal{H}_{\text{D}} &= \frac{\beta_{0}}{\hbar}(p_{x}\sigma_{x} - p_{y}\sigma_{y}).
    \end{split}
\end{equation}
We denote the effective mass as $m^{*}$, the chemical potential as $\mu$, the Rashba SOC strength as $\alpha_{0}$, the Dresselhaus SOC strength as $\beta_{0}$, and the exchange coupling strength as $J$. The spatial form of a general magnetic texture is given in spherical coordinates as $\textbf{m}=(\sin(\theta)\cos(\phi), \sin(\theta)\sin(\phi), \cos(\theta))$ where $\theta(x,y), \phi(x,y)$ defines the azimuthal and polar angles at $(x,y)$, respectively.

\par For further mathematical simplicity we can reformulate the intrinsic SOC within a real-space gauge field framework by rewriting the Hamiltonian as \cite{10.1063/1.3665219}
\begin{equation}
    \mathcal{H} = \frac{1}{2 m^{*}}\Bigl(\textbf{p} + e\textbf{A}\Bigr)^{2} - \mu' + J \textbf{m}\cdot\boldsymbol{\sigma},
    \label{continuum real-space gauge hamiltonian}
\end{equation}
where the gauge field $\textbf{A}=\textbf{A}_{R}+\textbf{A}_{D}$ is made up of the individual Rashba and Dresselhaus components,
\begin{equation}
    \begin{split}
        \textbf{A}_{\text{R}} &= \frac{\alpha_{0}m^{*}}{e \hbar}(-\sigma_{y},\sigma_{x},0),\\[5pt]
        \textbf{A}_{\text{D}} &= \frac{\beta_{0}m^{*}}{e \hbar}(-\sigma_{x},\sigma_{y},0).
    \end{split}
\end{equation}
Here we have gauge fields made of off-diagonal components which describe non-adiabatic processes which do not necessarily preserve the spin state of itinerant electrons. These processes tend to be of most importance in the regimes of weak exchange coupling and/or sharp magnetic textures\cite{PhysRevLett.117.027202, PhysRevB.98.195439, doi:10.7566/JPSJ.87.033705, PhysRevB.99.174425}. In performing this reformulation, there is also a chemical potential shift that we absorb in as $\mu'=\mu+\frac{e^{2}}{2 m^{*}}$. \\
\par It has been well established that itinerant electrons tend to realign their spins with a smooth, slowly varying magnetic texture. This effect gives rise to the phenomenon of emergent electrodynamics which has been studied for a variety of systems \cite{PhysRevLett.98.246601, PhysRevLett.102.186602, Schulz2012-fw, Nakabayashi_2014}. To investigate the interaction between the spin textures and SOC we will be investigating this emergent phenomenon through the use of the unitary spin-alignment transformation (SAT)\,\cite{PhysRevB.100.064504}
\begin{equation}
    U = \cos\biggl(\frac{\theta}{2}\biggr)\sigma_{0} + i\big(\sin(\phi)\sigma_{x} - \cos(\phi)\sigma_{y}\big)\sin\biggl(\frac{\theta}{2}\biggr).
    \label{SAT}
\end{equation}
The SAT represents a rotation about the $\textbf{n}=\frac{\textbf{m}\times\hat{z}}{|\textbf{m}\times\hat{z}|}$ axis in order to locally rotate $\textbf{m}(\textbf{r})$ to the positive $z$ direction. By applying the SAT to Eq.\,\eqref{continuum-hamiltonian} as $\mathcal{H}\to U^{\dagger} \mathcal{H} U$ we get the transformed Hamiltonian
\begin{equation}
    \mathcal{H}' = \frac{1}{2 m^{*}}\Bigl(\textbf{p} + e\textbf{A}'\Bigr)^{2} - \mu' + J\sigma_{z}.
\end{equation}
Under this transformation, there are new gauge fields induced along with changes to the original SOC gauge fields which represent the emergent electrodynamics. This new non-Abelian gauge field is defined as $\textbf{A}' = \textbf{A}'_{\text{s}} + \textbf{A}'_{\text{R}} + \textbf{A}'_{\text{D}}$ where
\begin{equation}
    \begin{split}
        \textbf{A}'_{\text{s}} &= \frac{-i\hbar}{e}U^{\dagger}\nabla U,\\[5pt]
        \textbf{A}'_{\text{R}} &= -U^{\dagger} \textbf{A}_{\text{R}} U,\\[5pt]
        \textbf{A}'_{\text{D}} &= -U^{\dagger}\textbf{A}_{\text{D}} U.
        \label{spin-rashba-dresselhaus gauge fields}
    \end{split}
\end{equation}
The $\textbf{A}'_{\text{s}}$ describes the SOC and chemical potential shift arising from the magnetic texture, while $\textbf{A}'_{\text{R}}$ and $\textbf{A}'_{\text{D}}$ describe the interaction between the spin texture and the intrinsic SOC.

The gauge field can be split into an adiabatic and non-adiabatic term as $\textbf{A}' = \textbf{A}'_{\text{Ad}} + \textbf{A}'_{\text{nAd}}$, where $\textbf{A}'_{\text{Ad}}$ contains the diagonal components of $\textbf{A}'$, while $\textbf{A}'_{\text{nAd}}$ contains the off-diagonal components. This allows the transformed Hamiltonian to be re-expressed as
\begin{equation}
    \mathcal{H}' = \frac{1}{2 m^{*}}\Bigl(\textbf{p} + e\textbf{A}'_{\text{Ad}}\Bigr)^{2} + \mathcal{H}'_{\text{SOC}} + \mathcal{H}'_{\mu} + J\sigma_{z}.
\end{equation}
From this we can define a new induced SOC arising from both the spin texture itself along with the interactions between the spin texture and the intrinsic Rashba and Dresselhaus SOC,
\begin{equation}
    \begin{split}
    \mathcal{H}'_{\text{SOC}}&=\frac{e}{2 m^{*}}\Bigl(\textbf{p}\cdot\textbf{A}'_{\text{nAd}} + \textbf{A}'_{\text{nAd}}\cdot\textbf{p}\Bigr)\\[5pt]
    &=\frac{p_{x}}{\hbar}(\alpha_{x}\sigma_{y} + \beta_{x}\sigma_{x}) - \frac{p_{y}}{\hbar}(\alpha_{y}\sigma_{x} + \beta_{y}\sigma_{y})\ .
    \label{Induced SOC}
    \end{split}
\end{equation}
Now we have an induced SOC that is spatially varying and potentially of different magnitudes in the $p_{x}$ and $p_{y}$ directions. Redefining the Hamiltonian in terms of non-adiabatic gauge fields also gives a spatially varying onsite chemical potential defined as
\begin{equation}
    \mathcal{H}'_{\mu}=\frac{e^{2}}{2 m^{*}}\Bigl(\textbf{A}'_{\text{Ad}}\cdot\textbf{A}'_{\text{nAd}} + \textbf{A}'_{\text{nAd}}\cdot\textbf{A}'_{\text{Ad}} + \textbf{A}'_{\text{nAd}}\cdot\textbf{A}'_{\text{nAd}}\Bigr) + \mu'.
\end{equation}
Note, however, that another equivalent way of representing this information is through the use of a non-adiabatic magnetic field defined as
\begin{equation}
    B_{i} = -\frac{1}{2}\epsilon_{ijk}F_{jk}.
\end{equation}
Here $F_{jk}$ represents the weakly coupled Yang-Mills curvature \cite{PhysRev.96.191} and can be defined as
\begin{equation}
    F_{jk} = \partial_{j}A_{k} - \partial_{k}A_{j} - \frac{i e}{\hbar}[A_{k}, A_{j}].
\end{equation}
Due to the symmetry of the curvature and the fact that we are using two-dimensional (2D) time-independent spin textures, the only non-zero Yang-Mills curvatures are $F_{xy}$ and $F_{yx}$. Thus, the magnetic field is defined by
\begin{equation}
    B_{z} = -\frac{1}{2}(F_{xy} - F_{yx}).
\end{equation}
By separating $B_{z}$, which is a $2\times2$ matrix, into Pauli components and only considering the $\sigma_{z}$ part, we can finally get the adiabatic magnetic field, which represents a time-reversal invariant spin-dependent magnetic field (i.e. $B_{\uparrow}=-B_{\downarrow}$). Mathematically, this is expressed as
\begin{equation}
    B_{z}^{\text{Ad}} = -\partial_{x}A_{y}^{\text{Ad}} + \partial_{y}A_{x}^{\text{Ad}} + \frac{i e}{\hbar}\Bigl[A_{y}^{\text{Ad}}, A_{x}^{\text{Ad}}\Bigr].
    \label{adiabatic magnetic field}
\end{equation}

\subsection{Skyrmion Continuum Results \label{Section IIB}}

\begin{figure}[t!]
    \centering
    \includegraphics[scale=1.3, keepaspectratio]{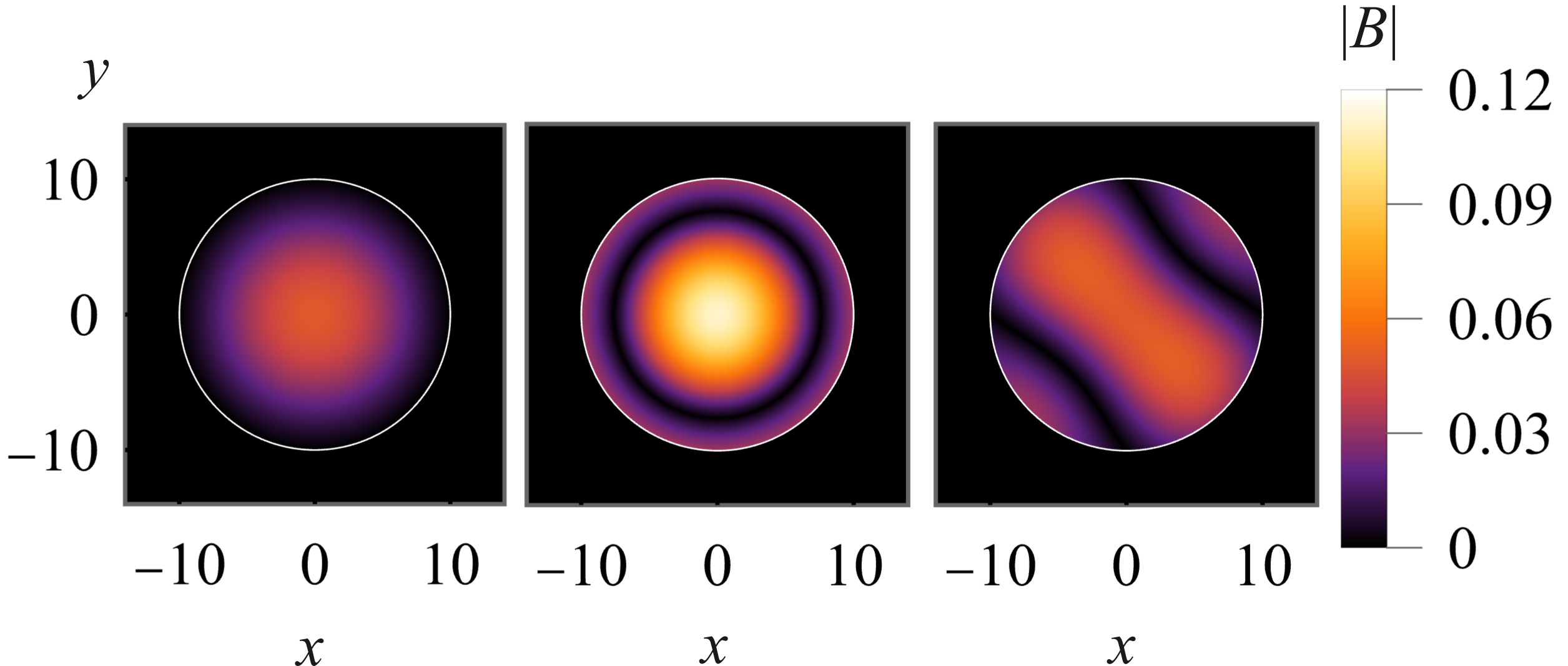}
    \caption{Induced magnetic field of a $R=10 a_{0}$ N\'eel skyrmion on a $40 a_0 \times 40 a_{0}$ square system for different intrinsic SOC strengths. Left: $\lambda_{\text{R}}=\lambda_{\text{D}}=0$. Center: $\lambda_{\text{R}}=0.2$\,eV,  $\lambda_{\text{D}}=0$. Right: $\lambda_{\text{R}}=0$, $\lambda_{\text{D}}=0.2$\,eV. Note that both $\lambda_{R,D}$ are related to the continuum SOC parameters via $\lambda_{R}=\frac{\alpha_{0}}{a_{0}}$ and $\lambda_{D}=\frac{\beta_{0}}{a_{0}}$ respectively.}
    \label{Graph2}
\end{figure}

    In order to better understand the interplay between intrinsic SOC and magnetic spin textures we consider a simplified skyrmion model,
\begin{equation}
    \begin{split}
        \theta(x,y) &= p\pi\text{ min}\Bigg(\frac{\sqrt{x^{2} + y^{2}}}{R}, 1\Bigg),\\[5pt]
        \phi(x,y) &= q \ \text{arctan}\Big(\frac{y}{x}\Big) + \phi_{0},
        \label{Neel Skyrmion}
    \end{split}
\end{equation}
where $R$ is the radius of the skyrmion. The radial and azimuthal winding numbers are defined as $p$ and $q$ respectively and $\phi_{0}$ represents the helicity of the skyrmion. For this Section we will be considering a N\'eel skyrmion defined as $(p,q,\phi_{0})=(1,1,0)$ of radius $R=10 a_{0}$, where $a_0$ is the lattice constant, in a $+\hat{z}$ out-of-plane ferromagnetic substrate with intrinsic Rashba and Dresselhaus SOC with amplitudes $\alpha_{0}$ and $\beta_{0}$, respectively. 

    Using the N\'eel skyrmion texture defined in Eq.\,\eqref{Neel Skyrmion}, we can use the results shown in Section\,\ref{Section IIA} to find expressions for the induced adiabatic magnetic field and SOC. The analytic expressions are provided in Appendix A. To compare these results with those based on the tight-binding model (see Section\,\ref{Section III}), we discretize the continuum system to obtain a relation between the continuum parameters and the tight-binding parameters, $\lambda_{R}=\frac{\alpha_{0}}{a_{0}}$ and $\lambda_{D}=\frac{\beta_{0}}{a_{0}}$. The analytic expressions for the conversion between continuum and tight-binding parameters are provided in Appendix B. A key assumption with the conversion is that the skyrmion texture is smooth and slowly varying with respect to the lattice.
    
    Using Eq.\,\eqref{adiabatic magnetic field} we calculate the adiabatic magnetic field as shown in Fig.\,\ref{Graph2} for varying intrinsic SOC strengths. For the case of no intrinsic SOC, as shown in Fig.\,\ref{Graph2}\,a, one can see that within the region of the skyrmion there exists an azimuthally symmetric magnetic field that peaks at the center of the skyrmion. Outside the skyrmion the adiabatic magnetic field is exactly zero due to the SAT being the identity matrix in the case of out-of-plane ferromagnetism. 
    
    In the case of a finite intrinsic Rashba SOC, $\lambda_{\text{R}}=0.2 \text{ eV}$, as shown in Fig.\,\ref{Graph2}\,b, the central region of the adiabatic magnetic field becomes amplified in strength. Further, the minimum of the magnetic field is no longer located at the edge of the skyrmion but now slightly inside of it. This can be understood as coming from the toy model we used for the skyrmion having a discontinuous first derivative with respect to the polar radial coordinate, $r$, at the boundary between the edge of the skyrmion and the ferromagnetic background,\cite{PhysRevB.95.064426}. 
    
    Finally, when considering the case of an intrinsic Dresselhaus SOC, $\lambda_{\text{D}}=0.2 \text{ eV}$, we find that the azimuthal symmetry of the adiabatic magnetic field is broken. This can be understood as coming from Dresselhaus SOC breaking azimuthal symmetry itself, and thus this property is passed on to the induced adiabatic magnetic field.

\begin{figure}[h!]
    \centering
    \includegraphics[scale=1.05, keepaspectratio]{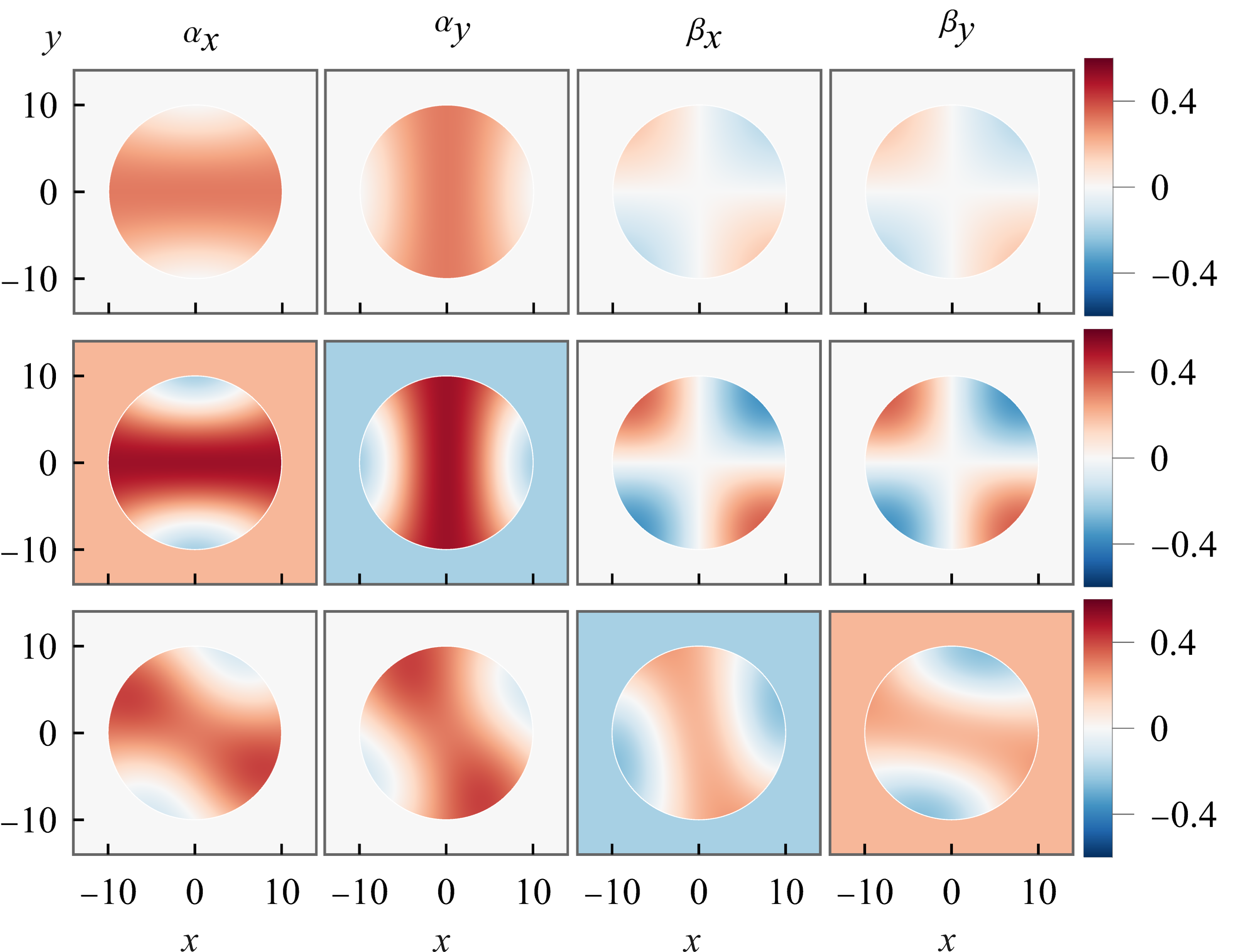}
    \caption{Induced Rashba and Dresselhaus SOC strengths in the $p_{x}$ and $p_{y}$ directions of a $R=10 a_{0}$ N\'eel skyrmion on a $40 a_0 \times 40 a_{0}$ square system for different intrinsic SOC strengths. Top Row: No SOC. Middle Row: $\lambda_{\text{R}}=0.2 \text{ eV}$. Bottom Row: $\lambda_{\text{D}}=0.2 \text{ eV}$. Note that both $\lambda_{R,D}$ are related to the continuum SOC parameters via $\lambda_{R}=\frac{\alpha_{0}}{a_{0}}$ and $\lambda_{D}=\frac{\beta_{0}}{a_{0}}$ respectively.}
    \label{Graph1}
\end{figure}

    Similar to before, we are also interested in investigating the induced SOC that comes from the interaction between intrinsic SOC and magnetic textures using 
    Eq.\,\eqref{Induced SOC}. The analytical expressions for the induced SOC are provided in Appendix B.  Specifically, we are interested in exploring how the addition of intrinsic SOC to a system with a magnetic texture leads to non-additive induced SOC effects.

    For the case of no intrinsic SOC, as shown in the top row of Fig.\,\ref{Graph1}, we recover the fact \cite{PhysRevB.85.020503, PhysRevB.82.045127, PhysRevB.102.224501} that a spin texture itself induces Rashba and Dresselhaus SOC under the SAT. Specifically, for a N\'eel skyrmion the induced Rashba SOC largely peaks along the axis of hopping while the induced Dresselhaus SOC peaks near the edges of the skyrmion. \par
    When we include an intrinsic Rashba SOC to the system, we find that there is non-trivial induced Rashba and Dresselhaus SOC being created. From the middle plot of Fig.\,\ref{Graph1}, we can see that the addition of intrinsic Rashba SOC leads to both the induced Rashba and Dresselhaus SOC to be amplified in magnitude in contrast to the naive expectation of intrinsic Rashba only affecting the induced Rashba's magnitude. We stress that within the skyrmion's region in Fig.\,\ref{Graph1} there are areas such that the induced SOC is zero, while the substrate region without the skyrmion shows constant intrinsic SOC. In these areas of vanishing induced SOC within the skyrmion, the intrinsic and induced SOC do not add; instead there is a cancellation between the two effects.

    Similarly, Fig.\,\ref{Graph1}\,c shows how the addition of intrinsic Dresselhaus SOC also leads to non-additive effects. However, we also see that the induced Rashba SOC becomes slightly rotated along with the shape of the induced Dresselhaus SOC becoming similar to the induced Rashba SOC. The rotation in SOC can be understood as coming from the breaking of azimuthal symmetry by the intrinsic Dresselhaus SOC. 
    
    In this Section we have applied a unitary transformation to the continuum Hamiltonian Eq.\,\eqref{continuum-hamiltonian} in order to get a new transformed Hamiltonian with induced adiabatic magnetic fields and SOC. The existence of the induced adiabatic magnetic field has been previously discussed \cite{PhysRevApplied.12.054032} along with proposals for experimental applications. However, we wish to assert that the calculated induced SOC are also observable quantities as we will discuss in the next Section.

%
%
\section{Local Spin Current \label{Section III}}


\subsection{Method \label{Section IIIA}}

    We can further corroborate the analytical Yang-Mills predictions for induced SOC with numerical simulations of a local current marker of induced SOC. This is achieved using the quantum transport software package KWANT\,\cite{Groth_2014} to simulate the coherent transport of itinerant electrons traversing a single N\'eel skyrmion. The tight-binding Hamiltonian we use is
\begin{equation}
    H = \sum_{\textbf{i}}c_{\textbf{i}}^{\dagger}(\mu + J\textbf{m}_{\textbf{i}}\cdot\boldsymbol{\sigma})c_{\textbf{i}}  - \sum_{\langle\textbf{i},\textbf{j}\rangle}c_{\textbf{i}}^{\dagger}t_{\textbf{i}\textbf{j}}c_{\textbf{j}},
    \label{tightbinding-hamiltonian}
\end{equation}
where $\mu$ is the onsite chemical potential which we assume to be constant, and $J$ the exchange coupling strength between the background spin texture $\textbf{m}_{\textbf{i}}$ at site $\textbf{i}$ and the itinerant electron with spin $\boldsymbol{\sigma}$. $c_{\textbf{i}}^{\dagger}=(c^{\dagger}_{\textbf{i}\uparrow}, c^{\dagger}_{\textbf{i}\downarrow})$ is a spinor containing usual fermionic creation operators at some site $\textbf{i}$ with $\uparrow,\downarrow$ referring to the spin projection along the quantization $z$ axis. $t_{\textbf{i}\textbf{j}}$ describes hopping between nearest-neighbor sites which incorporates intrinsic SOC as
\begin{equation}
    t_{\textbf{i}\textbf{j}} = \begin{cases}
        t +i\lambda_{\text{R}}\sigma_{y} + i\lambda_{\text{D}}\sigma_{x}, \qquad \textbf{j}=\textbf{i}\pm(a_{0},0),\\[5pt]
        t - i\lambda_{\text{R}}\sigma_{x} - i\lambda_{\text{D}}\sigma_{y}, \qquad \textbf{j}=\textbf{i}\pm(0,a_{0}),
    \end{cases}
\end{equation}
where $t$ is the hopping amplitude, $a_{0}$ is the lattice spacing constant, and $\lambda_{\text{R}}=\frac{\alpha_{0}}{a_{0}}, \lambda_{\text{D}}=\frac{\beta_{0}}{a_{0}}$ are the tight-binding SOC strengths. Note that this direct correspondence between the continuum Hamiltonian Eq.\,\ref{continuum-hamiltonian} and the tight-binding Hamiltonian Eq.\,\ref{tightbinding-hamiltonian} holds for low band filling\,\cite{PhysRevLett.95.046601}. We consider a system of an isolated N\'eel skyrmion of radius $R=10 a_{0}$ sitting on a square lattice of size $(N_{x},N_{y})=(40,40)$ with two semi-infinite ferromagnetic leads attached to the left and right edges of the system.

    Using KWANT, we are able to find the scattering states, $\psi$, originating from each of the leads. This can then be used to find the spin current\,\cite{RevModPhys.82.1539, An2012-da} for a single scattering state
\begin{equation}
    J_{\textbf{i}\textbf{j}}=i\Big[\psi^{\dagger}_{\textbf{j}}(H_{\textbf{i}\textbf{j}})^{\dagger} \textbf{M} \psi_{\textbf{i}} - \psi_{\textbf{i}}^{\dagger} \textbf{M} H_{\textbf{i}\textbf{j}} \psi_{\textbf{j}}\Big],
\end{equation}
which represents a spin flowing from lattice site $\textbf{j}$ to $\textbf{i}$ polarized in the $\textbf{M}=(\sigma_{x},\sigma_{y},\sigma_{z})$ direction. By summing over all the scattering states available, we get an observable total bond current between every nearest-neighbor site. 

    Note that it is possible to represent the intrinsic SOC in the continuum model terms of $\mathbf{k}$-dependent magnetic fields $\mathcal{H}_{\text{R}} = \textbf{B}_{\text{R}}(\textbf{k})\cdot\boldsymbol{\sigma}$ and $\mathcal{H}_{\text{D}} = \textbf{B}_{\text{D}}(\textbf{k})\cdot\boldsymbol{\sigma}$ where
\begin{equation}
    \begin{split}
        \textbf{B}_{\text{R}}(\textbf{k})&=\alpha_{0}(\textbf{k}\times\hat{z}),\\[5pt]
        \textbf{B}_{\text{D}}(\textbf{k})&=\beta_{0}(-k_{x}\hat{x} + k_{y}\hat{y}).
    \end{split}
\end{equation}
    Itinerant electrons on spin-orbit coupled materials tend to align with these Rashba and Dresselhaus fields, known as spin-momentum locking\,\cite{Manchon2015-ds}. 
    
    We can use spin-momentum locking as motivation for developing local markers for the effective SOC which we will label as the Rashba and Dresselhaus currents. For a tight-binding model, if we have a spin hopping in the $+\hat{x}$ direction, then spin-momentum locking would dictate that the spin axis would point in the $+\hat{y}$ direction in the presence of Rashba SOC and in the $+\hat{x}$ for Dresselhaus SOC. Similarly for a spin hopping in the $+\hat{y}$ direction the spin would align in the $-\hat{x}$ direction for Rashba SOC and the $-\hat{y}$ direction for Dresselhaus SOC. Returning to the two-terminal quantum transport simulation we can represent these ideas as calculating the spin-projected currents along different hopping directions as
\begin{equation}
    \begin{split}
        J_{\textbf{i}\textbf{j}}^{\text{R}} &= \begin{cases}
            i\Big[\psi_{\textbf{j}}^{\dagger}(H_{\textbf{i}\textbf{j}})^{\dagger}\sigma_{y}\psi_{\textbf{i}} - \psi_{\textbf{i}}^{\dagger}\sigma_{y}H_{\textbf{i}\textbf{j}}\psi_{\textbf{j}}\Big], ~ \textbf{j}=\textbf{i}\pm(a_{0},0)\\[8pt]
            i\Big[-\psi_{\textbf{j}}^{\dagger}(H_{\textbf{i}\textbf{j}})^{\dagger}\sigma_{x}\psi_{\textbf{i}} + \psi_{\textbf{i}}^{\dagger}\sigma_{x}H_{\textbf{i}\textbf{j}}\psi_{\textbf{j}}\Big], ~ \textbf{j}=\textbf{i}\pm(0,a_{0})
        \end{cases}\\[10pt]
        J_{\textbf{i}\textbf{j}}^{\text{D}} &= \begin{cases}
            i\Big[\psi_{\textbf{j}}^{\dagger}(H_{\textbf{i}\textbf{j}})^{\dagger}\sigma_{x}\psi_{\textbf{i}} - \psi_{\textbf{i}}^{\dagger}\sigma_{x}H_{\textbf{i}\textbf{j}}\psi_{\textbf{j}}\Big], ~ \textbf{j}=\textbf{i}\pm(a_{0},0)\\[8pt]
            i\Big[-\psi_{\textbf{j}}^{\dagger}(H_{\textbf{i}\textbf{j}})^{\dagger}\sigma_{y}\psi_{\textbf{i}} + \psi_{\textbf{i}}^{\dagger}\sigma_{y}H_{\textbf{i}\textbf{j}}\psi_{\textbf{j}}\Big], ~ \textbf{j}=\textbf{i}\pm(0,a_{0})
        \end{cases}
        \label{Rashba/Dresselhaus Current}
    \end{split}
\end{equation}
where $J_{\textbf{i}\textbf{j}}^{\text{R}},J_{\textbf{i}\textbf{j}}^{\text{D}}$ are the Rashba and Dresselhaus currents, respectively, and represent local markers for the Rashba and Dresselhaus SOC. However, realistically \cite{PhysRevB.95.064426}, it is necessary to consider contributions from each transverse mode $\psi_{\textbf{i}}$, originating form both the left and right leads. In this paper we sum all of these contributions to find the total Rashba and Dresselhaus currents. 

    So far we have used the Hamiltonian Eq.\,\eqref{tightbinding-hamiltonian} to calculate the Rashba and Dresselhaus currents in the laboratory frame. However, to corroborate with the results in Section\,\ref{Section II} it is necessary to apply the tight-binding version of the SAT Eq.\,\eqref{SAT} to the tight-binding Hamiltonian Eq.\,\eqref{tightbinding-hamiltonian} in order to get the system into the magnetic frame. The SAT at each lattice site $\textbf{r}$ is defined as
    \begin{equation}
        U_{\textbf{r}} = \exp\Big(i \frac{\theta}{2}\textbf{n}(\textbf{r})\cdot\boldsymbol{\sigma}\Big)\ .
        \label{tb sat}
    \end{equation}
    Using this transformed Hamiltonian we are then able to calculate the Rashba and Dresselhaus currents for comparison with the Yang-Mills results from Section\,\ref{Section II}.


\subsection{Skyrmion Effective Spin-orbit Coupling\label{Section IIIB}}
\begin{figure}[h!]
    \centering
    \includegraphics[scale=1.29, keepaspectratio]{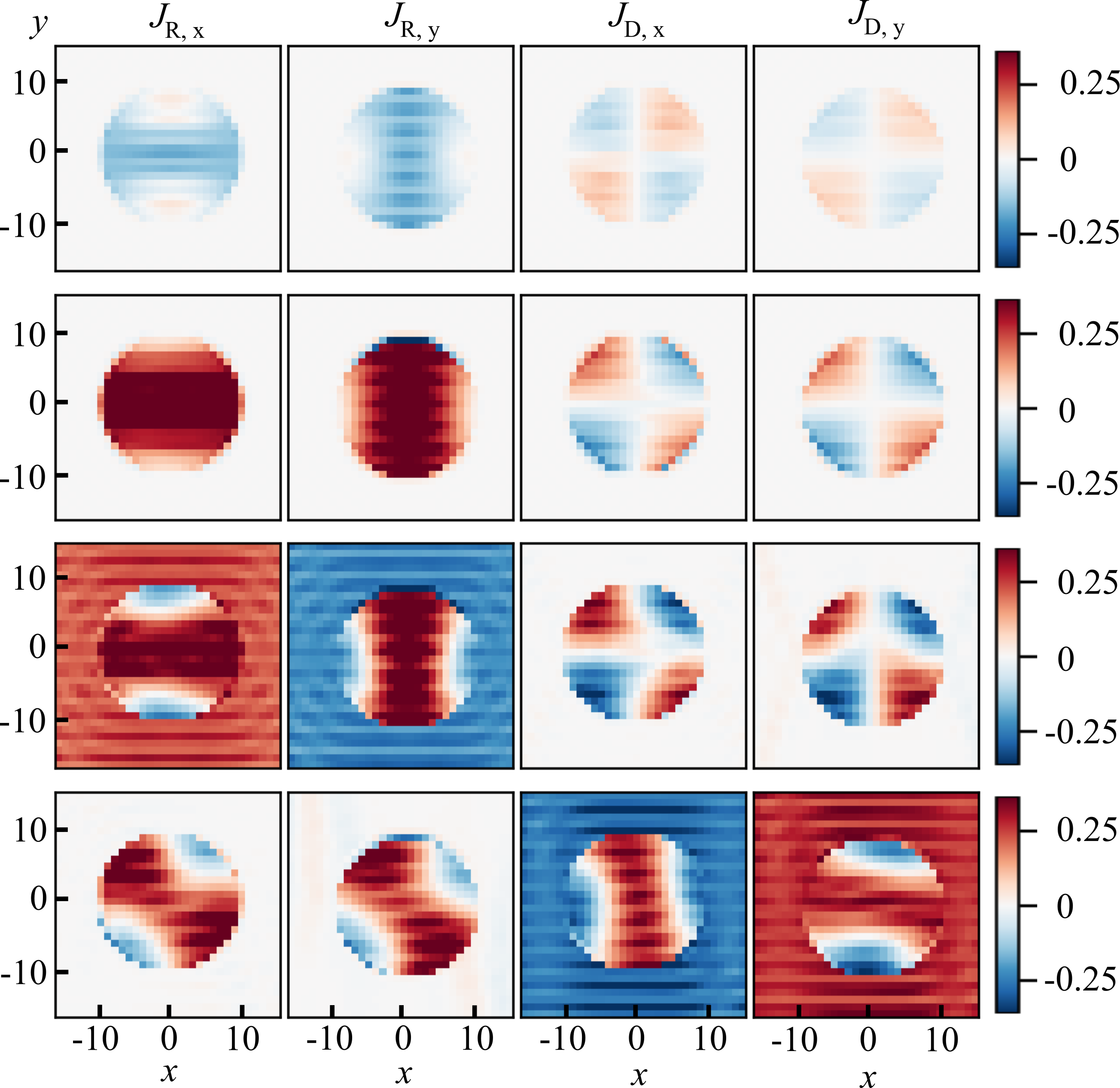}
    \caption{Induced Rashba and Dresselhaus SOC markers for different intrinsic SOC strengths. First row: No SOC in the laboratory frame. Second row: No SOC in the magnetic frame. Third row: $\lambda_{\text{R}}=0.2 \text{ eV}$ in the magnetic frame. Fourth row: $\lambda_{\text{D}}=0.2 \text{ eV}$ in the magnetic frame.}
    \label{Graph3}
\end{figure}

    We begin our comparison of the analytical results from Section\,\ref{Section II} with the numerical results Section\,\ref{Section III} by considering the case of a N\'eel skyrmion with no intrinsic SOC in the laboratory frame as shown in the 1st row of Fig.\,\ref{Graph3}. In this case we see the Rashba and Dresselhaus currents as measured from the laboratory perspective whilst noting the similarity in shape compared with the analytical results in the first row of Fig.\,\ref{Graph1}. 
    
    If we instead apply the SAT to Eq.\,\eqref{tightbinding-hamiltonian} before performing these simulations we can see the Rashba and Dresselhaus currents in the magnetic frame as shown in the second row of Fig.\,\ref{Graph3}. This transformed space represents a rotating frame with the spin quantization axis along magnetic texture such that itinerant electrons would see a background of uniform ferromagnetism coupled to the spin gauge fields of Section\,\ref{Section II}. These SOC markers show remarkable agreement with the analytical results in the first row of Fig.\,\ref{Graph1} up to a certain resolution coming from the discretized nature of the tight-binding model. Thus corroborating our previous analytical results along with providing a new numerical marker for investigating induced SOC. 
    
    As we add intrinsic Rashba and Dresselhaus SOC to the system, as shown in the third row of Fig.\,\ref{Graph3} and the fourth row of Fig.\,\ref{Graph3}, respectively, we again find notable agreement between analytical calculations and numerical simulations. In both cases the background SOC shows a rippling effect not present in the analytical results which comes from the nature of the tight-binding system being finite in the $+\hat{y}$ direction leading to finite-size effects. A secondary point to note is that the magnitude of the Rashba and Dresselhaus current markers also seem to reflect the difference in strength compared to analytical results. An example being how intrinsic Rashba SOC seems to scale the induced SOC more than intrinsic Dresselhaus SOC does.
    
    Experimentally, the likely path towards measuring SOC currents is through precise measurement of the texture of the skyrmion and the local spin-polarized current. It has been shown \cite{Birch2020-rj} that through a combination of scanning transmission X-ray microscopy and Lorentz transmission electron microscopy it is possible to measure the out-of-plane and in-plane components of the magnetic skyrmion textures. As these measurements become precise enough to measure at the lattice level, it should be possible to experimentally define a spin-alignment transformation using Eq.\,\eqref{tb sat}. Measurement of the local spin-polarized current is much more difficult due to the necessity of making spin-selective measurements on the atomic level. Previous literature has shown \cite{PhysRevLett.119.137202} that through the use of a four-probe spin-polarized scanning transmission microscopy experiment it is possible to measure the spin-dependent chemical potential and thus determine the spin polarization of the current. However, similar to before, current experimental techniques for spin current measurements are limited in precision to the $\mu$m scale. If this precision could be reduced to the atomic bond level, then it should be possible to experimentally verify the laboratory frame results of the first row of Fig.\,\ref{Graph3}. The direct measurement of spin currents in the magnetic frame is currently unexplored, however, a possible way around this could be the measurement of the local density of states through an scanning tunneling microscopy experiment to determine the local wavefunction combined with the experimentally found spin-alignment transformation previously discussed. Then, by applying a local frame transformation to Eq.\,\eqref{Rashba/Dresselhaus Current} this should give an expression for the SOC currents in the magnetic frame in terms of experimentally determined variables. The measurement of SOC currents not only corroborates the results of Sections\,\ref{Section II} and \ref{Section III} but also provides a new mechanism for classifying magnetic textures.


%
%
\section{Topological Hall Effect \label{Section IV}}

\subsection{Method \label{Section IVA}}

\begin{figure}[h!]
    \centering
    \includegraphics[scale=0.3, keepaspectratio]{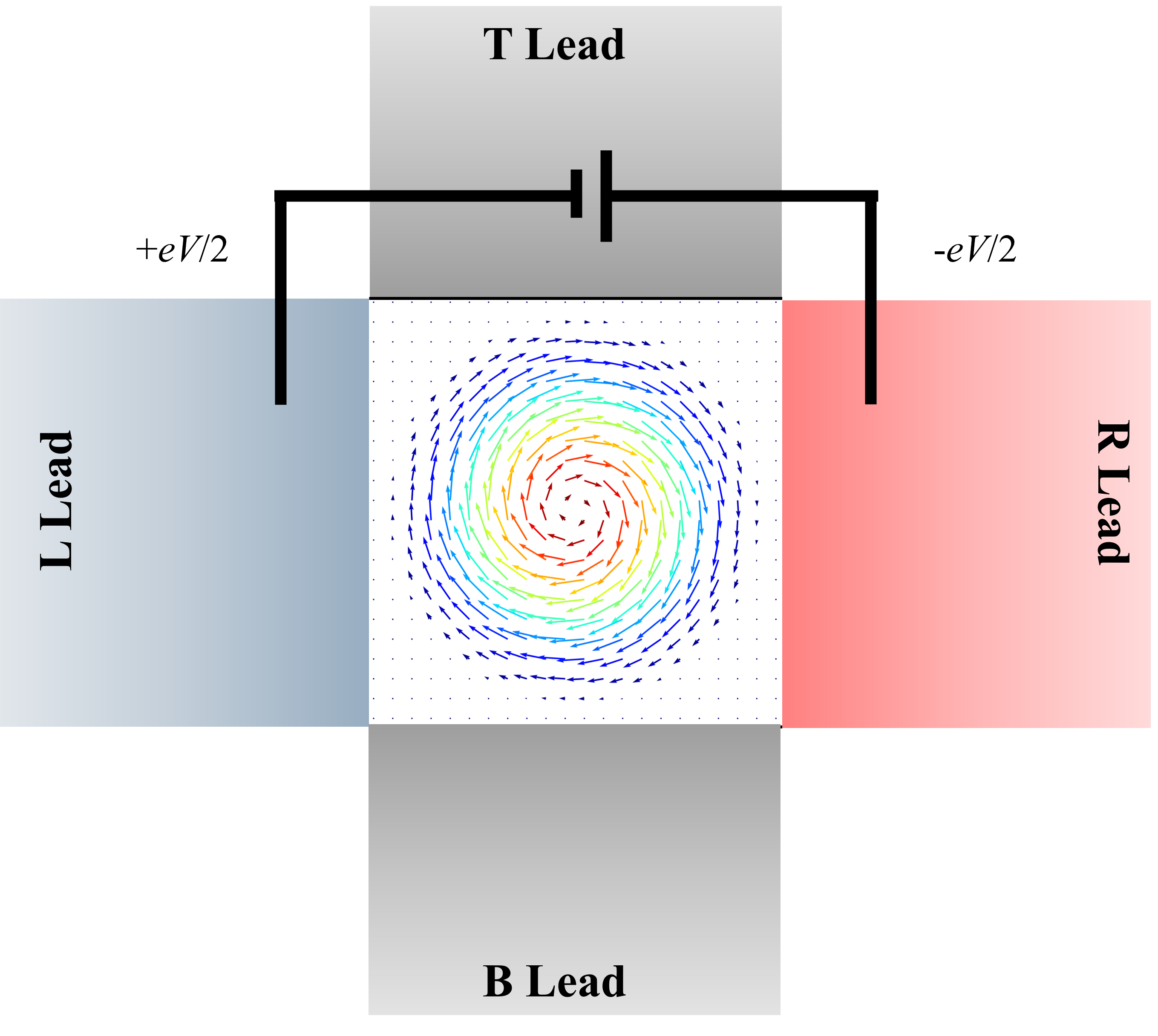}
    \caption{Schematic diagram of a four-terminal Hall experiment with a scattering region containing a magnetic Bloch skyrmion, in the presence of SOC, attached to four ferromagnetic leads: left (L), top (T), right (R), and bottom (B). With the system experiencing a longitudinal voltage bias of eV inducing a current to flow in from the L lead and get scattered thus giving a Hall response.}
    \label{fig:SKYRMIONLEAD}
\end{figure}

    In order to get the entire picture behind the interaction between spin textures and intrinsic SOC we also consider the topological Hall effect (THE) as a global transport measurement of this interaction. 
    
    As before, we use the same Hamiltonian Eq.\,\eqref{tightbinding-hamiltonian}, however, this time we consider a four terminal setup with a voltage bias on the left and right leads as shown in Fig.\,\ref{fig:SKYRMIONLEAD}.
    
    We employ the Landauer-B\"uttiker formalism\,\cite{PhysRevLett.57.1761} to investigate the coherent charge transport and scattering from a system with a longitudinal voltage bias. The terminal current of lead $\mu$ ($\mu=L, R, T, B$) is defined as
\begin{equation}
    I_{\mu} = \frac{e^{2}}{2\pi\hbar}\sum_{\mu\neq\nu}(T_{\nu\mu}V_{\mu} - T_{\mu\nu}V_{\nu}),
\end{equation}
where $V_{\mu}$ is the voltage at lead $\mu$ and $T_{\nu\mu}$ is the transmission coefficient for electrons from lead $\mu$ to lead $\nu$. 

    We calculate the terminal voltages following\,\cite{PhysRevB.95.064426} which can then be used to quantify the THE via the topological Hall angle
\begin{equation}
    \theta_{\text{TH}} = \frac{V_{\text{T}}-V_{\text{B}}}{V_{\text{R}}-V_{\text{L}}}.
\end{equation}
We consider systems with a chemical potential of $\mu=0$ on a $(N_{x}, N_{y})=(40,40)$ square lattice with a N\'eel skyrmion of radius $R=10 a_{0}$ located in the center. For the results in Fig.\,\ref{Graph5} we assume the strong exchange limit of $J=5 t$ in which the induced adiabatic magnetic fields are the dominant effect. 

We consider four different types of skyrmions in this section, being the N\'eel skyrmion and antiskyrmion which are defined by the parameters $(p,q,\phi_{0})=(1,\pm1,0)$, respectively. Further, we also consider the Bloch skyrmion and antiskyrmion defined by the parameters $(p,q,\phi_{0})=(1,\pm1,\frac{\pi}{2})$, respectively.



\subsection{Skyrmion Topological Hall Effect \label{Section IVB}}

\begin{figure}[h!]
    \centering
    \includegraphics[scale=1.0]{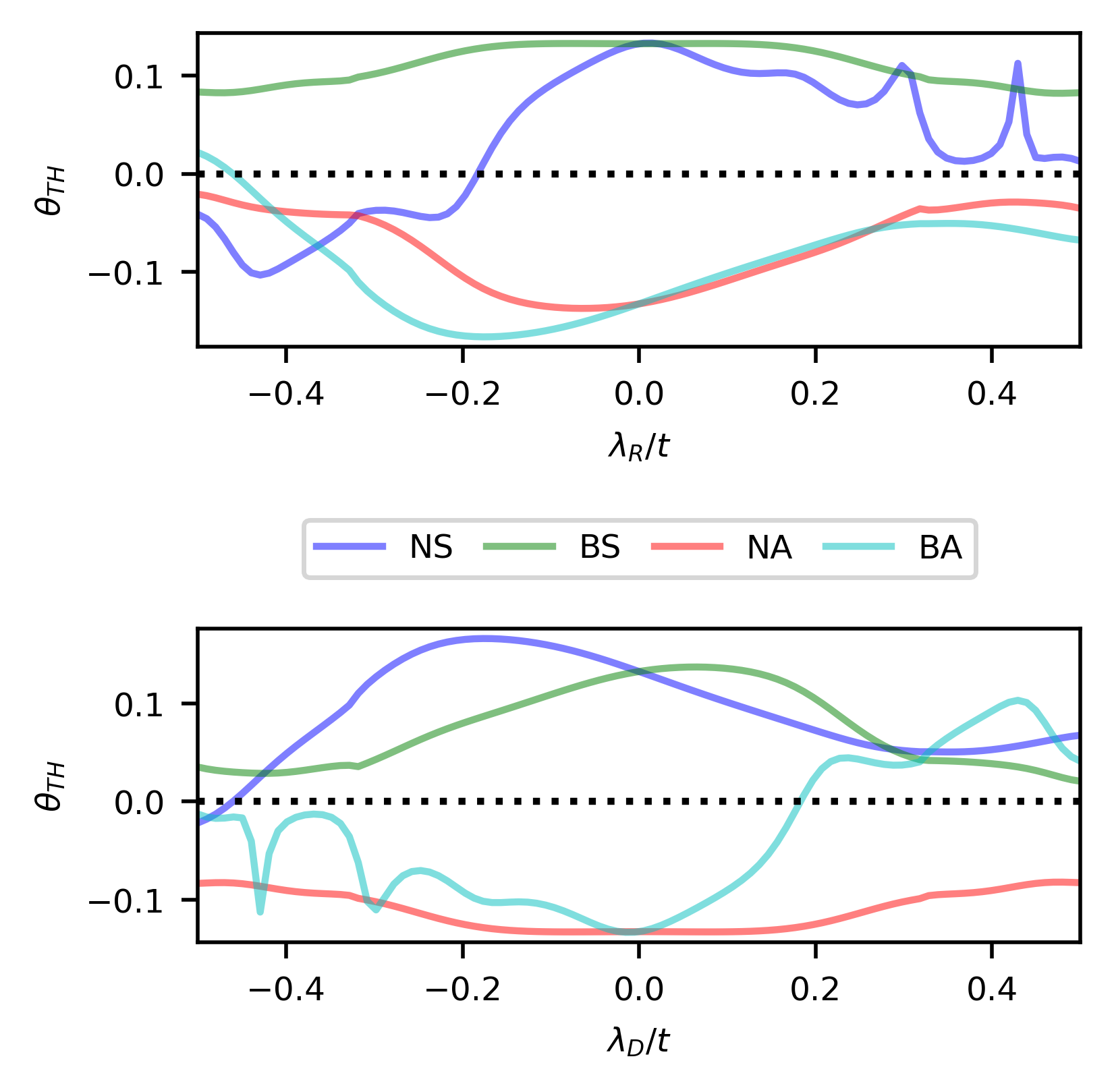}
    \caption{Topological Hall effect for varying SOC with different skyrmions. Top: Rashba Profile. Bottom: Dresselhaus profile.  NS: N\'eel skyrmion. BS: Bloch skyrmion. NA: N\'eel antiskyrmion. BA: Bloch antiskyrmion.} 
    \label{Graph5}
\end{figure}

    To gain a greater insight into the global transport properties relating to the interactions between skyrmions and intrinsic SOC we consider the Hall scattering from the system in Section\,\ref{Section IVA} at the band edge. For the chosen parameter set the system is at the band edge when the scattering regions fermi energy is $\epsilon_{F}=-8.8 t$. As shown in Fig.\,\ref{Graph5} we consider how itinerant electrons scatter off N\'eel skyrmions/antiskyrmions and Bloch skyrmions/antiskyrmions with respect to varying intrinsic Rashba SOC and Dresselhaus SOC. 
    
    For the case where the intrinsic SOC is zero, N\'eel and Bloch skyrmions scatter itinerant electrons at the same topological Hall angle whilst N\'eel and Bloch antiskyrmions scatter with same magnitude but in the opposite direction. This can be understood as the Hall angle for a non-spin-orbit-coupled system being dependent on the topological charge\,\cite{PhysRevB.75.012408}, $Q$, of the spin textures being considered. 
    
    In the presence of intrinsic SOC the skyrmion-antiskyrmion relation breaks down and each skyrmion type has a unique scattering profile which we will denote as the Rashba and Dresselhaus profiles. For a varying intrinsic Rashba SOC, as shown in Fig.\,\ref{Graph5}\,a, the N\'eel skyrmion has a zero crossing of the Hall angle, as predicted in Ref.\,\onlinecite{PhysRevApplied.12.054032}, whilst the Bloch skyrmion remains symmetric about zero SOC. Following this, the N\'eel and Bloch antiskyrmions both tend to have their peak Hall angles occurring for negative intrinsic Rashba SOC values. \par
    However, when we compare the Hall angle for varying Dresselhaus SOC, as shown in Fig.\,\ref{Graph5}\,b, we can see there is a clear relation between the scattering of skyrmions with varying Rashba and Dresselhaus SOC. Specifically, the N\'eel skyrmion Rashba scattering profile has the same shape as the Bloch antiskyrmion Dresselhaus scattering profile reflected about $\theta_{\text{TH}}=0$ and $\lambda_{D}=0$.
    \begin{figure}[h!]
        \centering
        \includegraphics[scale=0.47, keepaspectratio]{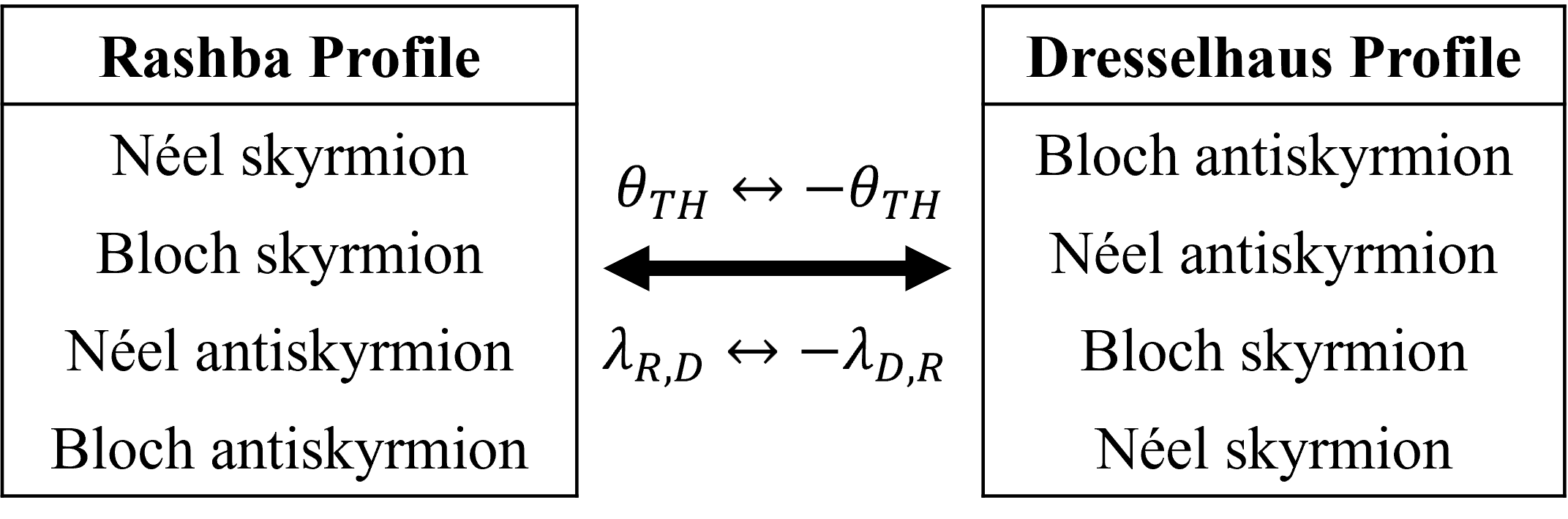}
        \caption{Relationship between the Rashba and Dresselhaus profiles. See main text for details.}
        \label{Table}
    \end{figure}
    
    Similar relations exist between other skyrmion pairs, as summarized by Fig.\,\ref{Table}\,, where we can see that the Rashba and Dresselhaus profiles are connected through a flipping of the sign of the azimuthal winding number, $q$, and a changing of the skyrmion type between N\'eel and Bloch through the helicity, $\phi_{0}$.
    
    Examining the structure of the Rashba and Dresselhaus profiles of Fig.\,\ref{Graph5} shows a nonlinear dependence on the intrinsic SOC. For example, the Bloch skyrmion Rashba profile, as shown in Fig.\,\ref{Graph5}\,a, only sees a relatively significant change at $\lambda_{\text{R}}\approx\pm0.2 \text{ eV}$ whereas the N\'eel skyrmion Rashba profile shows large swings in topological Hall angle throughout the entire parameter space. 
    
    We conclude this Section by noting that we are seeing the nonlinear behavior occurring at the system-wide quantum transport scale whereas the previous results demonstrated the nonlinearity at the local bond scale.

%
%
\section{Conclusion \label{Section V}}
    Magnetic skyrmions have emerged as key contenders for many spin-based technologies in the fields of spintronics and topological superconductivity. However, there lacks a domain-agnostic understanding of the interplay between magnetic skyrmions and SOC substrates. Previous literature has primarily focused on the Hall response of skyrmions coupled to SOC substrates whereas here we focus on the broader properties through a variety of theoretical techniques. 
    
    We began with the calculation of the induced adiabatic magnetic field and induced Rashba and Dresselhaus SOC coming from magnetic skyrmions grown on thin-film substrates possessing SOC using continuum model spintronic gauge theory. This is then corroborated through the introduction of a new induced SOC marker using spin-projected bond currents on a tight binding system. Finally, we investigate the mesoscopic scale properties of skyrmion-SOC substrate interactions using Landuer-B\"uttiker theory in order to connect the Rashba and Dresselhaus profiles of various skyrmion types. 
    



\begin{acknowledgments}
 S.R.\ acknowledges support from the Australian Research Council through Grant No.\ DP200101118 and DP240100168.
This research was undertaken using resources from the National Computational Infrastructure (NCI Australia), an NCRIS enabled capability supported by the Australian Government.
\end{acknowledgments}

\section*{Appendix A: General Skyrmion Spin Texture Results \label{Appendix A}}
Here we list the induced phenomena occurring from the application of the SAT to the continuum Hamiltonian for a N\'eel skyrmion. For mathematical simplicity we perform a coordinate change into polar coordinates: $(x,y)\to(r,\varphi)$.

    The induced adiabatic magnetic field has three components,
    \begin{equation}
        B_{z}^{\text{Ad}} = B_{z}^{\text{Ad, s}} + B_{z}^{\text{Ad, R}} + B_{z}^{\text{Ad, D}}.
        \label{mag}
    \end{equation}
    In the following equations we are referring to the magnetic flux quantum $\phi_{0}=\frac{h}{e}$.
    The components of $B_{z}^{\text{Ad}}$
    include a component, $B_{z}^{\text{Ad, s}}$, representing the adiabatic magnetic field arising from the spin texture itself
    \begin{equation}
        B_{z}^{\text{Ad, s}} = -\phi_{0}\frac{1}{4 r R}\sin\Big(\frac{\pi r}{R}\Big).
        \label{mag spin texture}
    \end{equation}
    There is also a component, $B_{z}^{\text{Ad, R}}$ describing the interaction between the spin texture and intrinsic Rashba SOC
    \begin{equation}
        B_{z}^{\text{Ad, R}} = -\phi_{0}\frac{\alpha_{0} m^{*}}{2 \pi r R \hbar^{2}}\Big[\pi r \cos\Big(\frac{\pi r}{R}\Big) + R\sin\Big(\frac{\pi r}{R}\Big)\Big].
        \label{mag rashba}
    \end{equation}
    Finally, a component, $B_{z}^{\text{Ad, D}}$, describing the interaction between the spin texture and intrinsic Dresselhaus SOC
    \begin{equation}
        B_{z}^{\text{Ad, D}} = \phi_{0}\frac{\beta_{0} m^{*}}{2 \pi r R \hbar^{2}}\Big[-\pi r \cos\Big(\frac{\pi r}{R}\Big) + R \sin\Big(\frac{\pi r}{R}\Big)\Big]\sin(2\varphi).
        \label{mag dresselhaus}
    \end{equation}
    
    The induced SOC can be written in the form
    \begin{equation}
        H'_{\text{SOC}} =  \hbar(\alpha_{x}\sigma_{y} + \beta_{x}\sigma_{x})p_{x} - \hbar(\alpha_{y}\sigma_{y} + \beta_{y}\sigma_{y})p_{y},
        \label{soc ham}
    \end{equation}
    where the induced Rashba SOC in the $p_{x}$ direction is given by
    \begin{equation}
        \begin{split}
            \alpha_{x} =& -\frac{\hbar^{2}}{2 r R m^{*}}\Big[\pi r \cos^{2}(\varphi) + R \sin\Big(\frac{\pi r}{R}\Big)\sin^{2}(\varphi)\Big]\\[5pt]
            &- \alpha_{0}\Big[\cos^{2}(\varphi) + \cos\Big(\frac{\pi r}{R}\Big)\sin^{2}(\varphi)\Big]\\[5pt]
            &+ \beta_{0}\sin^{2}\Big(\frac{\pi r}{2 R}\Big)\sin(2\varphi).
            \label{rashba x}
        \end{split}
    \end{equation}
    The induced Dresselhaus SOC in the $p_{x}$ direction is given by
    \begin{equation}
        \begin{split}
            \beta_{x} =& \frac{\hbar^{2}}{4 r R m^{*}}\Big[\pi r - R\sin\Big(\frac{\pi r}{R}\Big)\Big]\sin(2\varphi)\\[5pt]
            &+ \alpha_{0}\sin^{2}\Big(\frac{\pi r}{2 R}\Big)\sin(2\varphi)\\[5pt]
            &- \beta_{0}\Big[\cos\Big(\frac{\pi r}{R}\Big)\cos^{2}(\varphi) + \sin^{2}(2\varphi)\Big].
            \label{dresselhaus x}
        \end{split}
    \end{equation}
    The induced Rashba SOC in the $p_{y}$ direction is given by
    \begin{equation}
        \begin{split}
            \alpha_{y} =& \frac{\hbar^{2}}{2 r R m^{*}}\Big[R \cos^{2}(\varphi) \sin\Big(\frac{\pi r}{R}\Big) + \pi r \sin^{2}(\varphi)\Big]\\[5pt]
            &+ \alpha_{0}\Big[\cos\Big(\frac{\pi r}{R}\Big)\cos^{2}(\varphi) + \sin^{2}(\varphi)\Big]\\[5pt]
            &- 2\beta_{0}\cos(\varphi)\sin(\varphi)\sin^{2}\Big(\frac{\pi r}{2 R}\Big).
            \label{rashba y}
        \end{split}
    \end{equation}
    And finally the induced Dresselhaus SOC in the $p_{y}$ direction is given by
    \begin{equation}
        \begin{split}
            \beta_{y} &= \frac{\hbar^{2}}{2 r R m^{*}}\Big[-\pi r + R\sin\Big(\frac{\pi r}{R}\Big)\Big]\sin(\varphi)\\[5pt]
            &- 2\alpha_{0}\cos(\varphi)\sin(\varphi)\sin^{2}\Big(\frac{\pi r}{2 R}\Big)\\[5pt]
            &+ \beta_{0}\Big[\cos^{2}(\varphi) + \sin^{2}(\varphi)\cos\Big(\frac{\pi r}{R}\Big)\Big].
            \label{dresselhaus y}
        \end{split}
    \end{equation}

\section*{Appendix B: Equivalence of Continuum and Tight-Binding Parameters \label{Appendix B}}
    In order to compare the analytical results in Section\,\ref{Section II} with the numerical simulations in Section\,\ref{Section III} we need to ensure that both systems are equivalent in the limit that the skyrmion is large enough. Assuming a square lattice with lattice constant $a_{0}$, we refer to Ref.\,\onlinecite{PhysRevB.102.224501} to find the relations between continuum and tight-binding model parameters as
    \begin{equation}
        \begin{split}
            m^{*} &= \frac{\hbar^{2}}{2 t a_{0}^{2}},\\[5pt]
            \alpha_{0} &= a_{0} \lambda_{\text{R}},\\[5pt]
            \beta_{0} &= a_{0} \lambda_{\text{D}}.
        \end{split}
    \end{equation}
    These values can then be substituted into Eqs.\,\eqref{mag}-\eqref{dresselhaus y} in order to get the continuum model in terms of tight-binding parameters allowing the direct comparison between the results of Sections\,\ref{Section II} and \ref{Section III}. 
    
    Referring to Eq.\,\eqref{mag spin texture}, the component representing the adiabatic magnetic field arising from the spin texture becomes
    \begin{equation}
        B_{z}^{\text{Ad, s}} = -\phi_{0}\frac{1}{4 r R}\sin(\frac{\pi r}{R}).
    \end{equation}
    Referring to Eq.\,\eqref{mag rashba}, the component describing the interaction between the spin texture and intrinsic Rashba SOC becomes
    \begin{equation}
        B_{z}^{\text{Ad, R}} = -\phi_{0}\frac{\lambda_{\text{R}}}{4\pi r R t a_{0}}\Big[\pi r \cos\Big(\frac{\pi r}{R}\Big) + R\sin\Big(\frac{\pi r}{R}\Big)\Big],
    \end{equation}

    Finally, referring to Eq.\,\eqref{mag dresselhaus}, the component describing the interaction between the spin texture and intrinsic Dresselhaus SOC becomes
    \begin{equation}
        B_{z}^{\text{Ad, D}} = \phi_{0}\frac{\lambda_{\text{D}}}{4\pi r R t a_{0}}\Big[-\pi r \cos\Big(\frac{\pi r}{R}\Big) + R \sin\Big(\frac{\pi r}{R}\Big)\Big]\sin(2\varphi).
    \end{equation}

    Here the induced Rashba SOC in the $p_{x}$ direction is given by
    \begin{equation}
        \begin{split}
            \alpha_{x} =& -\frac{t a_{0}^{2}}{r R}\Big[\pi r \cos^{2}(\varphi) + R \sin\Big(\frac{\pi r}{R}\Big)\sin^{2}(\varphi)\Big]\\[5pt]
            &- a_{0}\lambda_{\text{R}}\Big[\cos^{2}(\varphi) + \cos\Big(\frac{\pi r}{R}\Big)\sin^{2}(\varphi)\Big]\\[5pt]
            &+ a_{0}\lambda_{\text{D}}\sin^{2}\Big(\frac{\pi r}{2 R}\Big)\sin(2\varphi).
        \end{split}
    \end{equation}

    The induced Dresselhaus SOC in the $p_{x}$ direction is given by
    \begin{equation}
        \begin{split}
            \beta_{x} =& \frac{t a_{0}^{2}}{2 r R}\Big[\pi r - R\sin\Big(\frac{\pi r}{R}\Big)\Big]\sin(2\varphi)\\[5pt]
            &+ a_{0}\lambda_{\text{R}}\sin^{2}\Big(\frac{\pi r}{2 R}\Big)\sin(2\varphi)\\[5pt]
            &- a_{0}\lambda_{\text{D}}\Big[\cos\Big(\frac{\pi r}{R}\Big)\cos^{2}(\varphi) + \sin^{2}(2\varphi)\Big].
        \end{split}
    \end{equation}
    
    The induced Rashba SOC in the $p_{y}$ direction is given by
    \begin{equation}
        \begin{split}
            \alpha_{y} =& \frac{t a_{0}^{2}}{r R}\Big[R \cos^{2}(\varphi) \sin\Big(\frac{\pi r}{R}\Big) + \pi r \sin^{2}(\varphi)\Big]\\[5pt]
            &+ a_{0}\lambda_{\text{R}}\Big[\cos\Big(\frac{\pi r}{R}\Big)\cos^{2}(\varphi) + \sin^{2}(\varphi)\Big]\\[5pt]
            &- 2a_{0}\lambda_{\text{D}}\cos(\varphi)\sin(\varphi)\sin^{2}\Big(\frac{\pi r}{2 R}\Big).
        \end{split}
    \end{equation}

    And finally the induced Dresselhaus SOC in the $p_{y}$ direction is given by
    \begin{equation}
        \begin{split}
            \beta_{y} =& \frac{t a_{0}^{2}}{r R}\Big[-\pi r + R\sin\Big(\frac{\pi r}{R}\Big)\Big]\sin(\varphi)\\[5pt]
            &- 2a_{0}\lambda_{\text{R}}\cos(\varphi)\sin(\varphi)\sin^{2}\Big(\frac{\pi r}{2 R}\Big)\\[5pt]
            &+ a_{0}\lambda_{\text{D}}\Big[\cos^{2}(\varphi) + \sin^{2}(\varphi)\cos\Big(\frac{\pi r}{R}\Big)\Big].
        \end{split}
    \end{equation}

\newpage
\phantom{x}

\bibliography{references}

\begin{thebibliography}{56}%
\makeatletter
\providecommand \@ifxundefined [1]{%
 \@ifx{#1\undefined}
}%
\providecommand \@ifnum [1]{%
 \ifnum #1\expandafter \@firstoftwo
 \else \expandafter \@secondoftwo
 \fi
}%
\providecommand \@ifx [1]{%
 \ifx #1\expandafter \@firstoftwo
 \else \expandafter \@secondoftwo
 \fi
}%
\providecommand \natexlab [1]{#1}%
\providecommand \enquote  [1]{``#1''}%
\providecommand \bibnamefont  [1]{#1}%
\providecommand \bibfnamefont [1]{#1}%
\providecommand \citenamefont [1]{#1}%
\providecommand \href@noop [0]{\@secondoftwo}%
\providecommand \href [0]{\begingroup \@sanitize@url \@href}%
\providecommand \@href[1]{\@@startlink{#1}\@@href}%
\providecommand \@@href[1]{\endgroup#1\@@endlink}%
\providecommand \@sanitize@url [0]{\catcode `\\12\catcode `\$12\catcode `\&12\catcode `\#12\catcode `\^12\catcode `\_12\catcode `\%12\relax}%
\providecommand \@@startlink[1]{}%
\providecommand \@@endlink[0]{}%
\providecommand \url  [0]{\begingroup\@sanitize@url \@url }%
\providecommand \@url [1]{\endgroup\@href {#1}{\urlprefix }}%
\providecommand \urlprefix  [0]{URL }%
\providecommand \Eprint [0]{\href }%
\providecommand \doibase [0]{https://doi.org/}%
\providecommand \selectlanguage [0]{\@gobble}%
\providecommand \bibinfo  [0]{\@secondoftwo}%
\providecommand \bibfield  [0]{\@secondoftwo}%
\providecommand \translation [1]{[#1]}%
\providecommand \BibitemOpen [0]{}%
\providecommand \bibitemStop [0]{}%
\providecommand \bibitemNoStop [0]{.\EOS\space}%
\providecommand \EOS [0]{\spacefactor3000\relax}%
\providecommand \BibitemShut  [1]{\csname bibitem#1\endcsname}%
\let\auto@bib@innerbib\@empty
\bibitem [{\citenamefont {\ifmmode \check{Z}\else \v{Z}\fi{}uti\ifmmode~\acute{c}\else \'{c}\fi{}}\ \emph {et~al.}(2004)\citenamefont {\ifmmode \check{Z}\else \v{Z}\fi{}uti\ifmmode~\acute{c}\else \'{c}\fi{}}, \citenamefont {Fabian},\ and\ \citenamefont {Das~Sarma}}]{RevModPhys.76.323}%
  \BibitemOpen
  \bibfield  {author} {\bibinfo {author} {\bibfnamefont {I.}~\bibnamefont {\ifmmode \check{Z}\else \v{Z}\fi{}uti\ifmmode~\acute{c}\else \'{c}\fi{}}}, \bibinfo {author} {\bibfnamefont {J.}~\bibnamefont {Fabian}},\ and\ \bibinfo {author} {\bibfnamefont {S.}~\bibnamefont {Das~Sarma}},\ }\href {https://doi.org/10.1103/RevModPhys.76.323} {\bibfield  {journal} {\bibinfo  {journal} {Rev. Mod. Phys.}\ }\textbf {\bibinfo {volume} {76}},\ \bibinfo {pages} {323} (\bibinfo {year} {2004})}\BibitemShut {NoStop}%
\bibitem [{\citenamefont {Hirohata}\ \emph {et~al.}(2020)\citenamefont {Hirohata}, \citenamefont {Yamada}, \citenamefont {Nakatani}, \citenamefont {Prejbeanu}, \citenamefont {Diény}, \citenamefont {Pirro},\ and\ \citenamefont {Hillebrands}}]{HIROHATA2020166711}%
  \BibitemOpen
  \bibfield  {author} {\bibinfo {author} {\bibfnamefont {A.}~\bibnamefont {Hirohata}}, \bibinfo {author} {\bibfnamefont {K.}~\bibnamefont {Yamada}}, \bibinfo {author} {\bibfnamefont {Y.}~\bibnamefont {Nakatani}}, \bibinfo {author} {\bibfnamefont {I.-L.}\ \bibnamefont {Prejbeanu}}, \bibinfo {author} {\bibfnamefont {B.}~\bibnamefont {Diény}}, \bibinfo {author} {\bibfnamefont {P.}~\bibnamefont {Pirro}},\ and\ \bibinfo {author} {\bibfnamefont {B.}~\bibnamefont {Hillebrands}},\ }\href {https://doi.org/https://doi.org/10.1016/j.jmmm.2020.166711} {\bibfield  {journal} {\bibinfo  {journal} {J. Magn. Magn. Mat}\ }\textbf {\bibinfo {volume} {509}},\ \bibinfo {pages} {166711} (\bibinfo {year} {2020})}\BibitemShut {NoStop}%
\bibitem [{\citenamefont {Wang}\ \emph {et~al.}(2022)\citenamefont {Wang}, \citenamefont {Bheemarasetty}, \citenamefont {Duan}, \citenamefont {Zhou},\ and\ \citenamefont {Xiao}}]{WANG2022169905}%
  \BibitemOpen
  \bibfield  {author} {\bibinfo {author} {\bibfnamefont {K.}~\bibnamefont {Wang}}, \bibinfo {author} {\bibfnamefont {V.}~\bibnamefont {Bheemarasetty}}, \bibinfo {author} {\bibfnamefont {J.}~\bibnamefont {Duan}}, \bibinfo {author} {\bibfnamefont {S.}~\bibnamefont {Zhou}},\ and\ \bibinfo {author} {\bibfnamefont {G.}~\bibnamefont {Xiao}},\ }\href {https://doi.org/https://doi.org/10.1016/j.jmmm.2022.169905} {\bibfield  {journal} {\bibinfo  {journal} {J. Magn. Magn. Mater.}\ }\textbf {\bibinfo {volume} {563}},\ \bibinfo {pages} {169905} (\bibinfo {year} {2022})}\BibitemShut {NoStop}%
\bibitem [{\citenamefont {Marrows}\ and\ \citenamefont {Zeissler}(2021)}]{10.1063/5.0072735}%
  \BibitemOpen
  \bibfield  {author} {\bibinfo {author} {\bibfnamefont {C.~H.}\ \bibnamefont {Marrows}}\ and\ \bibinfo {author} {\bibfnamefont {K.}~\bibnamefont {Zeissler}},\ }\href {https://doi.org/10.1063/5.0072735} {\bibfield  {journal} {\bibinfo  {journal} {Appl. Phys. Lett.}\ }\textbf {\bibinfo {volume} {119}},\ \bibinfo {pages} {250502} (\bibinfo {year} {2021})}\BibitemShut {NoStop}%
\bibitem [{\citenamefont {Song}\ \emph {et~al.}(2020)\citenamefont {Song}, \citenamefont {Jeong}, \citenamefont {Pan}, \citenamefont {Zhang}, \citenamefont {Xia}, \citenamefont {Cha}, \citenamefont {Park}, \citenamefont {Kim}, \citenamefont {Finizio}, \citenamefont {Raabe}, \citenamefont {Chang}, \citenamefont {Zhou}, \citenamefont {Zhao}, \citenamefont {Kang}, \citenamefont {Ju},\ and\ \citenamefont {Woo}}]{Song_2020}%
  \BibitemOpen
  \bibfield  {author} {\bibinfo {author} {\bibfnamefont {K.~M.}\ \bibnamefont {Song}}, \bibinfo {author} {\bibfnamefont {J.-S.}\ \bibnamefont {Jeong}}, \bibinfo {author} {\bibfnamefont {B.}~\bibnamefont {Pan}}, \bibinfo {author} {\bibfnamefont {X.}~\bibnamefont {Zhang}}, \bibinfo {author} {\bibfnamefont {J.}~\bibnamefont {Xia}}, \bibinfo {author} {\bibfnamefont {S.}~\bibnamefont {Cha}}, \bibinfo {author} {\bibfnamefont {T.-E.}\ \bibnamefont {Park}}, \bibinfo {author} {\bibfnamefont {K.}~\bibnamefont {Kim}}, \bibinfo {author} {\bibfnamefont {S.}~\bibnamefont {Finizio}}, \bibinfo {author} {\bibfnamefont {J.}~\bibnamefont {Raabe}}, \bibinfo {author} {\bibfnamefont {J.}~\bibnamefont {Chang}}, \bibinfo {author} {\bibfnamefont {Y.}~\bibnamefont {Zhou}}, \bibinfo {author} {\bibfnamefont {W.}~\bibnamefont {Zhao}}, \bibinfo {author} {\bibfnamefont {W.}~\bibnamefont {Kang}}, \bibinfo {author} {\bibfnamefont {H.}~\bibnamefont {Ju}},\ and\ \bibinfo {author} {\bibfnamefont {S.}~\bibnamefont {Woo}},\ }\href
  {https://doi.org/10.1038/s41928-020-0385-0} {\bibfield  {journal} {\bibinfo  {journal} {Nat. Electron.}\ }\textbf {\bibinfo {volume} {3}},\ \bibinfo {pages} {148–155} (\bibinfo {year} {2020})}\BibitemShut {NoStop}%
\bibitem [{\citenamefont {Mascot}\ \emph {et~al.}(2021)\citenamefont {Mascot}, \citenamefont {Bedow}, \citenamefont {Graham}, \citenamefont {Rachel},\ and\ \citenamefont {Morr}}]{Mascot2021-hz}%
  \BibitemOpen
  \bibfield  {author} {\bibinfo {author} {\bibfnamefont {E.}~\bibnamefont {Mascot}}, \bibinfo {author} {\bibfnamefont {J.}~\bibnamefont {Bedow}}, \bibinfo {author} {\bibfnamefont {M.}~\bibnamefont {Graham}}, \bibinfo {author} {\bibfnamefont {S.}~\bibnamefont {Rachel}},\ and\ \bibinfo {author} {\bibfnamefont {D.~K.}\ \bibnamefont {Morr}},\ }\href@noop {} {\bibfield  {journal} {\bibinfo  {journal} {npj Quantum Materials}\ }\textbf {\bibinfo {volume} {6}},\ \bibinfo {pages} {6} (\bibinfo {year} {2021})}\BibitemShut {NoStop}%
\bibitem [{\citenamefont {Brüning}\ \emph {et~al.}(2024)\citenamefont {Brüning}, \citenamefont {Bedow}, \citenamefont {Conte}, \citenamefont {von Bergmann}, \citenamefont {Morr},\ and\ \citenamefont {Wiesendanger}}]{brüning2024noncollinearpathtopologicalsuperconductivity}%
  \BibitemOpen
  \bibfield  {author} {\bibinfo {author} {\bibfnamefont {R.}~\bibnamefont {Brüning}}, \bibinfo {author} {\bibfnamefont {J.}~\bibnamefont {Bedow}}, \bibinfo {author} {\bibfnamefont {R.~L.}\ \bibnamefont {Conte}}, \bibinfo {author} {\bibfnamefont {K.}~\bibnamefont {von Bergmann}}, \bibinfo {author} {\bibfnamefont {D.~K.}\ \bibnamefont {Morr}},\ and\ \bibinfo {author} {\bibfnamefont {R.}~\bibnamefont {Wiesendanger}},\ }\href {https://arxiv.org/abs/2405.14673} {\bibinfo {title} {The non-collinear path to topological superconductivity}} (\bibinfo {year} {2024}),\ \Eprint {https://arxiv.org/abs/2405.14673} {arXiv:2405.14673} \BibitemShut {NoStop}%
\bibitem [{\citenamefont {Petrović}\ \emph {et~al.}(2024)\citenamefont {Petrović}, \citenamefont {Psaroudaki}, \citenamefont {Fischer}, \citenamefont {Garst},\ and\ \citenamefont {Panagopoulos}}]{petrović2024colloquiumquantumpropertiesfunctionalities}%
  \BibitemOpen
  \bibfield  {author} {\bibinfo {author} {\bibfnamefont {A.~P.}\ \bibnamefont {Petrović}}, \bibinfo {author} {\bibfnamefont {C.}~\bibnamefont {Psaroudaki}}, \bibinfo {author} {\bibfnamefont {P.}~\bibnamefont {Fischer}}, \bibinfo {author} {\bibfnamefont {M.}~\bibnamefont {Garst}},\ and\ \bibinfo {author} {\bibfnamefont {C.}~\bibnamefont {Panagopoulos}},\ }\href {https://arxiv.org/abs/2410.11427} {\bibinfo {title} {Colloquium: Quantum properties and functionalities of magnetic skyrmions}} (\bibinfo {year} {2024}),\ \Eprint {https://arxiv.org/abs/2410.11427} {arXiv:2410.11427} \BibitemShut {NoStop}%
\bibitem [{\citenamefont {Rex}\ \emph {et~al.}(2020)\citenamefont {Rex}, \citenamefont {Gornyi},\ and\ \citenamefont {Mirlin}}]{PhysRevB.102.224501}%
  \BibitemOpen
  \bibfield  {author} {\bibinfo {author} {\bibfnamefont {S.}~\bibnamefont {Rex}}, \bibinfo {author} {\bibfnamefont {I.~V.}\ \bibnamefont {Gornyi}},\ and\ \bibinfo {author} {\bibfnamefont {A.~D.}\ \bibnamefont {Mirlin}},\ }\href {https://doi.org/10.1103/PhysRevB.102.224501} {\bibfield  {journal} {\bibinfo  {journal} {Phys. Rev. B}\ }\textbf {\bibinfo {volume} {102}},\ \bibinfo {pages} {224501} (\bibinfo {year} {2020})}\BibitemShut {NoStop}%
\bibitem [{\citenamefont {Rex}\ \emph {et~al.}(2019)\citenamefont {Rex}, \citenamefont {Gornyi},\ and\ \citenamefont {Mirlin}}]{PhysRevB.100.064504}%
  \BibitemOpen
  \bibfield  {author} {\bibinfo {author} {\bibfnamefont {S.}~\bibnamefont {Rex}}, \bibinfo {author} {\bibfnamefont {I.~V.}\ \bibnamefont {Gornyi}},\ and\ \bibinfo {author} {\bibfnamefont {A.~D.}\ \bibnamefont {Mirlin}},\ }\href {https://doi.org/10.1103/PhysRevB.100.064504} {\bibfield  {journal} {\bibinfo  {journal} {Phys. Rev. B}\ }\textbf {\bibinfo {volume} {100}},\ \bibinfo {pages} {064504} (\bibinfo {year} {2019})}\BibitemShut {NoStop}%
\bibitem [{\citenamefont {Nothhelfer}\ \emph {et~al.}(2022)\citenamefont {Nothhelfer}, \citenamefont {D\'{\i}az}, \citenamefont {Kessler}, \citenamefont {Meng}, \citenamefont {Rizzi}, \citenamefont {Hals},\ and\ \citenamefont {Everschor-Sitte}}]{PhysRevB.105.224509}%
  \BibitemOpen
  \bibfield  {author} {\bibinfo {author} {\bibfnamefont {J.}~\bibnamefont {Nothhelfer}}, \bibinfo {author} {\bibfnamefont {S.~A.}\ \bibnamefont {D\'{\i}az}}, \bibinfo {author} {\bibfnamefont {S.}~\bibnamefont {Kessler}}, \bibinfo {author} {\bibfnamefont {T.}~\bibnamefont {Meng}}, \bibinfo {author} {\bibfnamefont {M.}~\bibnamefont {Rizzi}}, \bibinfo {author} {\bibfnamefont {K.~M.~D.}\ \bibnamefont {Hals}},\ and\ \bibinfo {author} {\bibfnamefont {K.}~\bibnamefont {Everschor-Sitte}},\ }\href {https://doi.org/10.1103/PhysRevB.105.224509} {\bibfield  {journal} {\bibinfo  {journal} {Phys. Rev. B}\ }\textbf {\bibinfo {volume} {105}},\ \bibinfo {pages} {224509} (\bibinfo {year} {2022})}\BibitemShut {NoStop}%
\bibitem [{\citenamefont {Lo~Conte}\ \emph {et~al.}(2024)\citenamefont {Lo~Conte}, \citenamefont {Wiebe}, \citenamefont {Rachel}, \citenamefont {Morr},\ and\ \citenamefont {Wiesendanger}}]{Lo_Conte2024-bs}%
  \BibitemOpen
  \bibfield  {author} {\bibinfo {author} {\bibfnamefont {R.}~\bibnamefont {Lo~Conte}}, \bibinfo {author} {\bibfnamefont {J.}~\bibnamefont {Wiebe}}, \bibinfo {author} {\bibfnamefont {S.}~\bibnamefont {Rachel}}, \bibinfo {author} {\bibfnamefont {D.~K.}\ \bibnamefont {Morr}},\ and\ \bibinfo {author} {\bibfnamefont {R.}~\bibnamefont {Wiesendanger}},\ }\href@noop {} {\bibfield  {journal} {\bibinfo  {journal} {Riv. Nuovo Cimento.}\ }\textbf {\bibinfo {volume} {47}},\ \bibinfo {pages} {453} (\bibinfo {year} {2024})}\BibitemShut {NoStop}%
\bibitem [{\citenamefont {Je}\ \emph {et~al.}(2020)\citenamefont {Je}, \citenamefont {Han}, \citenamefont {Kim}, \citenamefont {Montoya}, \citenamefont {Chao}, \citenamefont {Hong}, \citenamefont {Fullerton}, \citenamefont {Lee}, \citenamefont {Lee}, \citenamefont {Im},\ and\ \citenamefont {Hong}}]{Je2020-ze}%
  \BibitemOpen
  \bibfield  {author} {\bibinfo {author} {\bibfnamefont {S.-G.}\ \bibnamefont {Je}}, \bibinfo {author} {\bibfnamefont {H.-S.}\ \bibnamefont {Han}}, \bibinfo {author} {\bibfnamefont {S.~K.}\ \bibnamefont {Kim}}, \bibinfo {author} {\bibfnamefont {S.~A.}\ \bibnamefont {Montoya}}, \bibinfo {author} {\bibfnamefont {W.}~\bibnamefont {Chao}}, \bibinfo {author} {\bibfnamefont {I.-S.}\ \bibnamefont {Hong}}, \bibinfo {author} {\bibfnamefont {E.~E.}\ \bibnamefont {Fullerton}}, \bibinfo {author} {\bibfnamefont {K.-S.}\ \bibnamefont {Lee}}, \bibinfo {author} {\bibfnamefont {K.-J.}\ \bibnamefont {Lee}}, \bibinfo {author} {\bibfnamefont {M.-Y.}\ \bibnamefont {Im}},\ and\ \bibinfo {author} {\bibfnamefont {J.-I.}\ \bibnamefont {Hong}},\ }\href@noop {} {\bibfield  {journal} {\bibinfo  {journal} {ACS Nano}\ }\textbf {\bibinfo {volume} {14}},\ \bibinfo {pages} {3251} (\bibinfo {year} {2020})}\BibitemShut {NoStop}%
\bibitem [{\citenamefont {Sotnikov}\ \emph {et~al.}(2023)\citenamefont {Sotnikov}, \citenamefont {Stepanov}, \citenamefont {Katsnelson}, \citenamefont {Mila},\ and\ \citenamefont {Mazurenko}}]{PhysRevX.13.041027}%
  \BibitemOpen
  \bibfield  {author} {\bibinfo {author} {\bibfnamefont {O.~M.}\ \bibnamefont {Sotnikov}}, \bibinfo {author} {\bibfnamefont {E.~A.}\ \bibnamefont {Stepanov}}, \bibinfo {author} {\bibfnamefont {M.~I.}\ \bibnamefont {Katsnelson}}, \bibinfo {author} {\bibfnamefont {F.}~\bibnamefont {Mila}},\ and\ \bibinfo {author} {\bibfnamefont {V.~V.}\ \bibnamefont {Mazurenko}},\ }\href {https://doi.org/10.1103/PhysRevX.13.041027} {\bibfield  {journal} {\bibinfo  {journal} {Phys. Rev. X}\ }\textbf {\bibinfo {volume} {13}},\ \bibinfo {pages} {041027} (\bibinfo {year} {2023})}\BibitemShut {NoStop}%
\bibitem [{\citenamefont {Buhrandt}\ and\ \citenamefont {Fritz}(2013)}]{PhysRevB.88.195137}%
  \BibitemOpen
  \bibfield  {author} {\bibinfo {author} {\bibfnamefont {S.}~\bibnamefont {Buhrandt}}\ and\ \bibinfo {author} {\bibfnamefont {L.}~\bibnamefont {Fritz}},\ }\href {https://doi.org/10.1103/PhysRevB.88.195137} {\bibfield  {journal} {\bibinfo  {journal} {Phys. Rev. B}\ }\textbf {\bibinfo {volume} {88}},\ \bibinfo {pages} {195137} (\bibinfo {year} {2013})}\BibitemShut {NoStop}%
\bibitem [{\citenamefont {Laliena}\ and\ \citenamefont {Campo}(2017)}]{PhysRevB.96.134420}%
  \BibitemOpen
  \bibfield  {author} {\bibinfo {author} {\bibfnamefont {V.}~\bibnamefont {Laliena}}\ and\ \bibinfo {author} {\bibfnamefont {J.}~\bibnamefont {Campo}},\ }\href {https://doi.org/10.1103/PhysRevB.96.134420} {\bibfield  {journal} {\bibinfo  {journal} {Phys. Rev. B}\ }\textbf {\bibinfo {volume} {96}},\ \bibinfo {pages} {134420} (\bibinfo {year} {2017})}\BibitemShut {NoStop}%
\bibitem [{\citenamefont {G{\"o}bel}\ \emph {et~al.}(2021)\citenamefont {G{\"o}bel}, \citenamefont {Mertig},\ and\ \citenamefont {Tretiakov}}]{Gobel2021}%
  \BibitemOpen
  \bibfield  {author} {\bibinfo {author} {\bibfnamefont {B.}~\bibnamefont {G{\"o}bel}}, \bibinfo {author} {\bibfnamefont {I.}~\bibnamefont {Mertig}},\ and\ \bibinfo {author} {\bibfnamefont {O.~A.}\ \bibnamefont {Tretiakov}},\ }\href {https://doi.org/https://doi.org/10.1016/j.physrep.2020.10.001} {\bibfield  {journal} {\bibinfo  {journal} {Phys. Rep.}\ }\textbf {\bibinfo {volume} {895}},\ \bibinfo {pages} {1} (\bibinfo {year} {2021})}\BibitemShut {NoStop}%
\bibitem [{\citenamefont {Yang}\ \emph {et~al.}(2016)\citenamefont {Yang}, \citenamefont {Stano}, \citenamefont {Klinovaja},\ and\ \citenamefont {Loss}}]{PhysRevB.93.224505}%
  \BibitemOpen
  \bibfield  {author} {\bibinfo {author} {\bibfnamefont {G.}~\bibnamefont {Yang}}, \bibinfo {author} {\bibfnamefont {P.}~\bibnamefont {Stano}}, \bibinfo {author} {\bibfnamefont {J.}~\bibnamefont {Klinovaja}},\ and\ \bibinfo {author} {\bibfnamefont {D.}~\bibnamefont {Loss}},\ }\href {https://doi.org/10.1103/PhysRevB.93.224505} {\bibfield  {journal} {\bibinfo  {journal} {Phys. Rev. B}\ }\textbf {\bibinfo {volume} {93}},\ \bibinfo {pages} {224505} (\bibinfo {year} {2016})}\BibitemShut {NoStop}%
\bibitem [{\citenamefont {Wiesendanger}(2016)}]{Wiesendanger2016-gn}%
  \BibitemOpen
  \bibfield  {author} {\bibinfo {author} {\bibfnamefont {R.}~\bibnamefont {Wiesendanger}},\ }\href@noop {} {\bibfield  {journal} {\bibinfo  {journal} {Nat. Rev. Mat.}\ }\textbf {\bibinfo {volume} {1}},\ \bibinfo {pages} {16044} (\bibinfo {year} {2016})}\BibitemShut {NoStop}%
\bibitem [{\citenamefont {Zhao}\ \emph {et~al.}(2024)\citenamefont {Zhao}, \citenamefont {Hua}, \citenamefont {Song}, \citenamefont {Yu},\ and\ \citenamefont {Jiang}}]{ZHAO20242370}%
  \BibitemOpen
  \bibfield  {author} {\bibinfo {author} {\bibfnamefont {L.}~\bibnamefont {Zhao}}, \bibinfo {author} {\bibfnamefont {C.}~\bibnamefont {Hua}}, \bibinfo {author} {\bibfnamefont {C.}~\bibnamefont {Song}}, \bibinfo {author} {\bibfnamefont {W.}~\bibnamefont {Yu}},\ and\ \bibinfo {author} {\bibfnamefont {W.}~\bibnamefont {Jiang}},\ }\href {https://doi.org/https://doi.org/10.1016/j.scib.2024.05.035} {\bibfield  {journal} {\bibinfo  {journal} {Sci. Bull.}\ }\textbf {\bibinfo {volume} {69}},\ \bibinfo {pages} {2370} (\bibinfo {year} {2024})}\BibitemShut {NoStop}%
\bibitem [{\citenamefont {Volovik}(1987)}]{GEVolovik_1987}%
  \BibitemOpen
  \bibfield  {author} {\bibinfo {author} {\bibfnamefont {G.~E.}\ \bibnamefont {Volovik}},\ }\href {https://doi.org/10.1088/0022-3719/20/7/003} {\bibfield  {journal} {\bibinfo  {journal} {J. Phys. C: Solid State Phys.}\ }\textbf {\bibinfo {volume} {20}},\ \bibinfo {pages} {L83} (\bibinfo {year} {1987})}\BibitemShut {NoStop}%
\bibitem [{\citenamefont {Bazaliy}\ \emph {et~al.}(1998)\citenamefont {Bazaliy}, \citenamefont {Jones},\ and\ \citenamefont {Zhang}}]{PhysRevB.57.R3213}%
  \BibitemOpen
  \bibfield  {author} {\bibinfo {author} {\bibfnamefont {Y.~B.}\ \bibnamefont {Bazaliy}}, \bibinfo {author} {\bibfnamefont {B.~A.}\ \bibnamefont {Jones}},\ and\ \bibinfo {author} {\bibfnamefont {S.-C.}\ \bibnamefont {Zhang}},\ }\href {https://doi.org/10.1103/PhysRevB.57.R3213} {\bibfield  {journal} {\bibinfo  {journal} {Phys. Rev. B}\ }\textbf {\bibinfo {volume} {57}},\ \bibinfo {pages} {R3213} (\bibinfo {year} {1998})}\BibitemShut {NoStop}%
\bibitem [{\citenamefont {Barnes}\ and\ \citenamefont {Maekawa}(2007)}]{PhysRevLett.98.246601}%
  \BibitemOpen
  \bibfield  {author} {\bibinfo {author} {\bibfnamefont {S.~E.}\ \bibnamefont {Barnes}}\ and\ \bibinfo {author} {\bibfnamefont {S.}~\bibnamefont {Maekawa}},\ }\href {https://doi.org/10.1103/PhysRevLett.98.246601} {\bibfield  {journal} {\bibinfo  {journal} {Phys. Rev. Lett.}\ }\textbf {\bibinfo {volume} {98}},\ \bibinfo {pages} {246601} (\bibinfo {year} {2007})}\BibitemShut {NoStop}%
\bibitem [{\citenamefont {Zang}\ \emph {et~al.}(2011)\citenamefont {Zang}, \citenamefont {Mostovoy}, \citenamefont {Han},\ and\ \citenamefont {Nagaosa}}]{PhysRevLett.107.136804}%
  \BibitemOpen
  \bibfield  {author} {\bibinfo {author} {\bibfnamefont {J.}~\bibnamefont {Zang}}, \bibinfo {author} {\bibfnamefont {M.}~\bibnamefont {Mostovoy}}, \bibinfo {author} {\bibfnamefont {J.~H.}\ \bibnamefont {Han}},\ and\ \bibinfo {author} {\bibfnamefont {N.}~\bibnamefont {Nagaosa}},\ }\href {https://doi.org/10.1103/PhysRevLett.107.136804} {\bibfield  {journal} {\bibinfo  {journal} {Phys. Rev. Lett.}\ }\textbf {\bibinfo {volume} {107}},\ \bibinfo {pages} {136804} (\bibinfo {year} {2011})}\BibitemShut {NoStop}%
\bibitem [{\citenamefont {Schulz}\ \emph {et~al.}(2012{\natexlab{a}})\citenamefont {Schulz}, \citenamefont {Ritz}, \citenamefont {Bauer}, \citenamefont {Halder}, \citenamefont {Wagner}, \citenamefont {Franz}, \citenamefont {Pfleiderer}, \citenamefont {Everschor}, \citenamefont {Garst},\ and\ \citenamefont {Rosch}}]{Schulz2012-bz}%
  \BibitemOpen
  \bibfield  {author} {\bibinfo {author} {\bibfnamefont {T.}~\bibnamefont {Schulz}}, \bibinfo {author} {\bibfnamefont {R.}~\bibnamefont {Ritz}}, \bibinfo {author} {\bibfnamefont {A.}~\bibnamefont {Bauer}}, \bibinfo {author} {\bibfnamefont {M.}~\bibnamefont {Halder}}, \bibinfo {author} {\bibfnamefont {M.}~\bibnamefont {Wagner}}, \bibinfo {author} {\bibfnamefont {C.}~\bibnamefont {Franz}}, \bibinfo {author} {\bibfnamefont {C.}~\bibnamefont {Pfleiderer}}, \bibinfo {author} {\bibfnamefont {K.}~\bibnamefont {Everschor}}, \bibinfo {author} {\bibfnamefont {M.}~\bibnamefont {Garst}},\ and\ \bibinfo {author} {\bibfnamefont {A.}~\bibnamefont {Rosch}},\ }\href@noop {} {\bibfield  {journal} {\bibinfo  {journal} {Nat. Phys.}\ }\textbf {\bibinfo {volume} {8}},\ \bibinfo {pages} {301} (\bibinfo {year} {2012}{\natexlab{a}})}\BibitemShut {NoStop}%
\bibitem [{\citenamefont {Chen}\ and\ \citenamefont {Byrnes}(2019)}]{PhysRevB.99.184427}%
  \BibitemOpen
  \bibfield  {author} {\bibinfo {author} {\bibfnamefont {T.}~\bibnamefont {Chen}}\ and\ \bibinfo {author} {\bibfnamefont {T.}~\bibnamefont {Byrnes}},\ }\href {https://doi.org/10.1103/PhysRevB.99.184427} {\bibfield  {journal} {\bibinfo  {journal} {Phys. Rev. B}\ }\textbf {\bibinfo {volume} {99}},\ \bibinfo {pages} {184427} (\bibinfo {year} {2019})}\BibitemShut {NoStop}%
\bibitem [{\citenamefont {Akosa}\ \emph {et~al.}(2019)\citenamefont {Akosa}, \citenamefont {Li}, \citenamefont {Tatara},\ and\ \citenamefont {Tretiakov}}]{PhysRevApplied.12.054032}%
  \BibitemOpen
  \bibfield  {author} {\bibinfo {author} {\bibfnamefont {C.~A.}\ \bibnamefont {Akosa}}, \bibinfo {author} {\bibfnamefont {H.}~\bibnamefont {Li}}, \bibinfo {author} {\bibfnamefont {G.}~\bibnamefont {Tatara}},\ and\ \bibinfo {author} {\bibfnamefont {O.~A.}\ \bibnamefont {Tretiakov}},\ }\href {https://doi.org/10.1103/PhysRevApplied.12.054032} {\bibfield  {journal} {\bibinfo  {journal} {Phys. Rev. Appl.}\ }\textbf {\bibinfo {volume} {12}},\ \bibinfo {pages} {054032} (\bibinfo {year} {2019})}\BibitemShut {NoStop}%
\bibitem [{\citenamefont {Ndiaye}\ \emph {et~al.}(2017)\citenamefont {Ndiaye}, \citenamefont {Akosa},\ and\ \citenamefont {Manchon}}]{PhysRevB.95.064426}%
  \BibitemOpen
  \bibfield  {author} {\bibinfo {author} {\bibfnamefont {P.~B.}\ \bibnamefont {Ndiaye}}, \bibinfo {author} {\bibfnamefont {C.~A.}\ \bibnamefont {Akosa}},\ and\ \bibinfo {author} {\bibfnamefont {A.}~\bibnamefont {Manchon}},\ }\href {https://doi.org/10.1103/PhysRevB.95.064426} {\bibfield  {journal} {\bibinfo  {journal} {Phys. Rev. B}\ }\textbf {\bibinfo {volume} {95}},\ \bibinfo {pages} {064426} (\bibinfo {year} {2017})}\BibitemShut {NoStop}%
\bibitem [{\citenamefont {Yin}\ \emph {et~al.}(2015)\citenamefont {Yin}, \citenamefont {Liu}, \citenamefont {Barlas}, \citenamefont {Zang},\ and\ \citenamefont {Lake}}]{PhysRevB.92.024411}%
  \BibitemOpen
  \bibfield  {author} {\bibinfo {author} {\bibfnamefont {G.}~\bibnamefont {Yin}}, \bibinfo {author} {\bibfnamefont {Y.}~\bibnamefont {Liu}}, \bibinfo {author} {\bibfnamefont {Y.}~\bibnamefont {Barlas}}, \bibinfo {author} {\bibfnamefont {J.}~\bibnamefont {Zang}},\ and\ \bibinfo {author} {\bibfnamefont {R.~K.}\ \bibnamefont {Lake}},\ }\href {https://doi.org/10.1103/PhysRevB.92.024411} {\bibfield  {journal} {\bibinfo  {journal} {Phys. Rev. B}\ }\textbf {\bibinfo {volume} {92}},\ \bibinfo {pages} {024411} (\bibinfo {year} {2015})}\BibitemShut {NoStop}%
\bibitem [{\citenamefont {Tan}\ \emph {et~al.}(2020)\citenamefont {Tan}, \citenamefont {Chen}, \citenamefont {Ho}, \citenamefont {Huang}, \citenamefont {Jalil}, \citenamefont {Chang},\ and\ \citenamefont {Murakami}}]{TAN20201}%
  \BibitemOpen
  \bibfield  {author} {\bibinfo {author} {\bibfnamefont {S.~G.}\ \bibnamefont {Tan}}, \bibinfo {author} {\bibfnamefont {S.-H.}\ \bibnamefont {Chen}}, \bibinfo {author} {\bibfnamefont {C.~S.}\ \bibnamefont {Ho}}, \bibinfo {author} {\bibfnamefont {C.-C.}\ \bibnamefont {Huang}}, \bibinfo {author} {\bibfnamefont {M.~B.}\ \bibnamefont {Jalil}}, \bibinfo {author} {\bibfnamefont {C.~R.}\ \bibnamefont {Chang}},\ and\ \bibinfo {author} {\bibfnamefont {S.}~\bibnamefont {Murakami}},\ }\href {https://doi.org/https://doi.org/10.1016/j.physrep.2020.08.002} {\bibfield  {journal} {\bibinfo  {journal} {Phys. Rep.}\ }\textbf {\bibinfo {volume} {882}},\ \bibinfo {pages} {1} (\bibinfo {year} {2020})}\BibitemShut {NoStop}%
\bibitem [{\citenamefont {Shen}(2005)}]{PhysRevLett.95.187203}%
  \BibitemOpen
  \bibfield  {author} {\bibinfo {author} {\bibfnamefont {S.-Q.}\ \bibnamefont {Shen}},\ }\href {https://doi.org/10.1103/PhysRevLett.95.187203} {\bibfield  {journal} {\bibinfo  {journal} {Phys. Rev. Lett.}\ }\textbf {\bibinfo {volume} {95}},\ \bibinfo {pages} {187203} (\bibinfo {year} {2005})}\BibitemShut {NoStop}%
\bibitem [{\citenamefont {Mühlbauer}\ \emph {et~al.}(2009)\citenamefont {Mühlbauer}, \citenamefont {Binz}, \citenamefont {Jonietz}, \citenamefont {Pfleiderer}, \citenamefont {Rosch}, \citenamefont {Neubauer}, \citenamefont {Georgii},\ and\ \citenamefont {Böni}}]{doi:10.1126/science.1166767}%
  \BibitemOpen
  \bibfield  {author} {\bibinfo {author} {\bibfnamefont {S.}~\bibnamefont {Mühlbauer}}, \bibinfo {author} {\bibfnamefont {B.}~\bibnamefont {Binz}}, \bibinfo {author} {\bibfnamefont {F.}~\bibnamefont {Jonietz}}, \bibinfo {author} {\bibfnamefont {C.}~\bibnamefont {Pfleiderer}}, \bibinfo {author} {\bibfnamefont {A.}~\bibnamefont {Rosch}}, \bibinfo {author} {\bibfnamefont {A.}~\bibnamefont {Neubauer}}, \bibinfo {author} {\bibfnamefont {R.}~\bibnamefont {Georgii}},\ and\ \bibinfo {author} {\bibfnamefont {P.}~\bibnamefont {Böni}},\ }\href {https://doi.org/10.1126/science.1166767} {\bibfield  {journal} {\bibinfo  {journal} {Science}\ }\textbf {\bibinfo {volume} {323}},\ \bibinfo {pages} {915} (\bibinfo {year} {2009})}\BibitemShut {NoStop}%
\bibitem [{\citenamefont {Yu}\ \emph {et~al.}(2010)\citenamefont {Yu}, \citenamefont {Onose}, \citenamefont {Kanazawa}, \citenamefont {Park}, \citenamefont {Han}, \citenamefont {Matsui}, \citenamefont {Nagaosa},\ and\ \citenamefont {Tokura}}]{Yu2010-ca}%
  \BibitemOpen
  \bibfield  {author} {\bibinfo {author} {\bibfnamefont {X.~Z.}\ \bibnamefont {Yu}}, \bibinfo {author} {\bibfnamefont {Y.}~\bibnamefont {Onose}}, \bibinfo {author} {\bibfnamefont {N.}~\bibnamefont {Kanazawa}}, \bibinfo {author} {\bibfnamefont {J.~H.}\ \bibnamefont {Park}}, \bibinfo {author} {\bibfnamefont {J.~H.}\ \bibnamefont {Han}}, \bibinfo {author} {\bibfnamefont {Y.}~\bibnamefont {Matsui}}, \bibinfo {author} {\bibfnamefont {N.}~\bibnamefont {Nagaosa}},\ and\ \bibinfo {author} {\bibfnamefont {Y.}~\bibnamefont {Tokura}},\ }\href@noop {} {\bibfield  {journal} {\bibinfo  {journal} {Nat.}\ }\textbf {\bibinfo {volume} {465}},\ \bibinfo {pages} {901} (\bibinfo {year} {2010})}\BibitemShut {NoStop}%
\bibitem [{\citenamefont {Seki}\ \emph {et~al.}(2012)\citenamefont {Seki}, \citenamefont {Yu}, \citenamefont {Ishiwata},\ and\ \citenamefont {Tokura}}]{doi:10.1126/science.1214143}%
  \BibitemOpen
  \bibfield  {author} {\bibinfo {author} {\bibfnamefont {S.}~\bibnamefont {Seki}}, \bibinfo {author} {\bibfnamefont {X.~Z.}\ \bibnamefont {Yu}}, \bibinfo {author} {\bibfnamefont {S.}~\bibnamefont {Ishiwata}},\ and\ \bibinfo {author} {\bibfnamefont {Y.}~\bibnamefont {Tokura}},\ }\href {https://doi.org/10.1126/science.1214143} {\bibfield  {journal} {\bibinfo  {journal} {Science}\ }\textbf {\bibinfo {volume} {336}},\ \bibinfo {pages} {198} (\bibinfo {year} {2012})}\BibitemShut {NoStop}%
\bibitem [{\citenamefont {K{\'e}zsm{\'a}rki}\ \emph {et~al.}(2015)\citenamefont {K{\'e}zsm{\'a}rki}, \citenamefont {Bord{\'a}cs}, \citenamefont {Milde}, \citenamefont {Neuber}, \citenamefont {Eng}, \citenamefont {White}, \citenamefont {R{\o}nnow}, \citenamefont {Dewhurst}, \citenamefont {Mochizuki}, \citenamefont {Yanai}, \citenamefont {Nakamura}, \citenamefont {Ehlers}, \citenamefont {Tsurkan},\ and\ \citenamefont {Loidl}}]{Kezsmarki2015-mv}%
  \BibitemOpen
  \bibfield  {author} {\bibinfo {author} {\bibfnamefont {I.}~\bibnamefont {K{\'e}zsm{\'a}rki}}, \bibinfo {author} {\bibfnamefont {S.}~\bibnamefont {Bord{\'a}cs}}, \bibinfo {author} {\bibfnamefont {P.}~\bibnamefont {Milde}}, \bibinfo {author} {\bibfnamefont {E.}~\bibnamefont {Neuber}}, \bibinfo {author} {\bibfnamefont {L.~M.}\ \bibnamefont {Eng}}, \bibinfo {author} {\bibfnamefont {J.~S.}\ \bibnamefont {White}}, \bibinfo {author} {\bibfnamefont {H.~M.}\ \bibnamefont {R{\o}nnow}}, \bibinfo {author} {\bibfnamefont {C.~D.}\ \bibnamefont {Dewhurst}}, \bibinfo {author} {\bibfnamefont {M.}~\bibnamefont {Mochizuki}}, \bibinfo {author} {\bibfnamefont {K.}~\bibnamefont {Yanai}}, \bibinfo {author} {\bibfnamefont {H.}~\bibnamefont {Nakamura}}, \bibinfo {author} {\bibfnamefont {D.}~\bibnamefont {Ehlers}}, \bibinfo {author} {\bibfnamefont {V.}~\bibnamefont {Tsurkan}},\ and\ \bibinfo {author} {\bibfnamefont {A.}~\bibnamefont {Loidl}},\ }\href@noop {} {\bibfield  {journal} {\bibinfo  {journal} {Nat. Mat.}\ }\textbf {\bibinfo
  {volume} {14}},\ \bibinfo {pages} {1116} (\bibinfo {year} {2015})}\BibitemShut {NoStop}%
\bibitem [{\citenamefont {Kurumaji}\ \emph {et~al.}(2017)\citenamefont {Kurumaji}, \citenamefont {Nakajima}, \citenamefont {Ukleev}, \citenamefont {Feoktystov}, \citenamefont {Arima}, \citenamefont {Kakurai},\ and\ \citenamefont {Tokura}}]{PhysRevLett.119.237201}%
  \BibitemOpen
  \bibfield  {author} {\bibinfo {author} {\bibfnamefont {T.}~\bibnamefont {Kurumaji}}, \bibinfo {author} {\bibfnamefont {T.}~\bibnamefont {Nakajima}}, \bibinfo {author} {\bibfnamefont {V.}~\bibnamefont {Ukleev}}, \bibinfo {author} {\bibfnamefont {A.}~\bibnamefont {Feoktystov}}, \bibinfo {author} {\bibfnamefont {T.-h.}\ \bibnamefont {Arima}}, \bibinfo {author} {\bibfnamefont {K.}~\bibnamefont {Kakurai}},\ and\ \bibinfo {author} {\bibfnamefont {Y.}~\bibnamefont {Tokura}},\ }\href {https://doi.org/10.1103/PhysRevLett.119.237201} {\bibfield  {journal} {\bibinfo  {journal} {Phys. Rev. Lett.}\ }\textbf {\bibinfo {volume} {119}},\ \bibinfo {pages} {237201} (\bibinfo {year} {2017})}\BibitemShut {NoStop}%
\bibitem [{\citenamefont {Fujita}\ \emph {et~al.}(2011)\citenamefont {Fujita}, \citenamefont {Jalil}, \citenamefont {Tan},\ and\ \citenamefont {Murakami}}]{10.1063/1.3665219}%
  \BibitemOpen
  \bibfield  {author} {\bibinfo {author} {\bibfnamefont {T.}~\bibnamefont {Fujita}}, \bibinfo {author} {\bibfnamefont {M.~B.~A.}\ \bibnamefont {Jalil}}, \bibinfo {author} {\bibfnamefont {S.~G.}\ \bibnamefont {Tan}},\ and\ \bibinfo {author} {\bibfnamefont {S.}~\bibnamefont {Murakami}},\ }\href {https://doi.org/10.1063/1.3665219} {\bibfield  {journal} {\bibinfo  {journal} {J. Appl. Phys.}\ }\textbf {\bibinfo {volume} {110}},\ \bibinfo {pages} {121301} (\bibinfo {year} {2011})}\BibitemShut {NoStop}%
\bibitem [{\citenamefont {Denisov}\ \emph {et~al.}(2016)\citenamefont {Denisov}, \citenamefont {Rozhansky}, \citenamefont {Averkiev},\ and\ \citenamefont {L\"ahderanta}}]{PhysRevLett.117.027202}%
  \BibitemOpen
  \bibfield  {author} {\bibinfo {author} {\bibfnamefont {K.~S.}\ \bibnamefont {Denisov}}, \bibinfo {author} {\bibfnamefont {I.~V.}\ \bibnamefont {Rozhansky}}, \bibinfo {author} {\bibfnamefont {N.~S.}\ \bibnamefont {Averkiev}},\ and\ \bibinfo {author} {\bibfnamefont {E.}~\bibnamefont {L\"ahderanta}},\ }\href {https://doi.org/10.1103/PhysRevLett.117.027202} {\bibfield  {journal} {\bibinfo  {journal} {Phys. Rev. Lett.}\ }\textbf {\bibinfo {volume} {117}},\ \bibinfo {pages} {027202} (\bibinfo {year} {2016})}\BibitemShut {NoStop}%
\bibitem [{\citenamefont {Denisov}\ \emph {et~al.}(2018)\citenamefont {Denisov}, \citenamefont {Rozhansky}, \citenamefont {Averkiev},\ and\ \citenamefont {L\"ahderanta}}]{PhysRevB.98.195439}%
  \BibitemOpen
  \bibfield  {author} {\bibinfo {author} {\bibfnamefont {K.~S.}\ \bibnamefont {Denisov}}, \bibinfo {author} {\bibfnamefont {I.~V.}\ \bibnamefont {Rozhansky}}, \bibinfo {author} {\bibfnamefont {N.~S.}\ \bibnamefont {Averkiev}},\ and\ \bibinfo {author} {\bibfnamefont {E.}~\bibnamefont {L\"ahderanta}},\ }\href {https://doi.org/10.1103/PhysRevB.98.195439} {\bibfield  {journal} {\bibinfo  {journal} {Phys. Rev. B}\ }\textbf {\bibinfo {volume} {98}},\ \bibinfo {pages} {195439} (\bibinfo {year} {2018})}\BibitemShut {NoStop}%
\bibitem [{\citenamefont {Nakazawa}\ \emph {et~al.}(2018)\citenamefont {Nakazawa}, \citenamefont {Bibes},\ and\ \citenamefont {Kohno}}]{doi:10.7566/JPSJ.87.033705}%
  \BibitemOpen
  \bibfield  {author} {\bibinfo {author} {\bibfnamefont {K.}~\bibnamefont {Nakazawa}}, \bibinfo {author} {\bibfnamefont {M.}~\bibnamefont {Bibes}},\ and\ \bibinfo {author} {\bibfnamefont {H.}~\bibnamefont {Kohno}},\ }\href {https://doi.org/10.7566/JPSJ.87.033705} {\bibfield  {journal} {\bibinfo  {journal} {JPJS}\ }\textbf {\bibinfo {volume} {87}},\ \bibinfo {pages} {033705} (\bibinfo {year} {2018})}\BibitemShut {NoStop}%
\bibitem [{\citenamefont {Nakazawa}\ and\ \citenamefont {Kohno}(2019)}]{PhysRevB.99.174425}%
  \BibitemOpen
  \bibfield  {author} {\bibinfo {author} {\bibfnamefont {K.}~\bibnamefont {Nakazawa}}\ and\ \bibinfo {author} {\bibfnamefont {H.}~\bibnamefont {Kohno}},\ }\href {https://doi.org/10.1103/PhysRevB.99.174425} {\bibfield  {journal} {\bibinfo  {journal} {Phys. Rev. B}\ }\textbf {\bibinfo {volume} {99}},\ \bibinfo {pages} {174425} (\bibinfo {year} {2019})}\BibitemShut {NoStop}%
\bibitem [{\citenamefont {Neubauer}\ \emph {et~al.}(2009)\citenamefont {Neubauer}, \citenamefont {Pfleiderer}, \citenamefont {Binz}, \citenamefont {Rosch}, \citenamefont {Ritz}, \citenamefont {Niklowitz},\ and\ \citenamefont {B\"oni}}]{PhysRevLett.102.186602}%
  \BibitemOpen
  \bibfield  {author} {\bibinfo {author} {\bibfnamefont {A.}~\bibnamefont {Neubauer}}, \bibinfo {author} {\bibfnamefont {C.}~\bibnamefont {Pfleiderer}}, \bibinfo {author} {\bibfnamefont {B.}~\bibnamefont {Binz}}, \bibinfo {author} {\bibfnamefont {A.}~\bibnamefont {Rosch}}, \bibinfo {author} {\bibfnamefont {R.}~\bibnamefont {Ritz}}, \bibinfo {author} {\bibfnamefont {P.~G.}\ \bibnamefont {Niklowitz}},\ and\ \bibinfo {author} {\bibfnamefont {P.}~\bibnamefont {B\"oni}},\ }\href {https://doi.org/10.1103/PhysRevLett.102.186602} {\bibfield  {journal} {\bibinfo  {journal} {Phys. Rev. Lett.}\ }\textbf {\bibinfo {volume} {102}},\ \bibinfo {pages} {186602} (\bibinfo {year} {2009})}\BibitemShut {NoStop}%
\bibitem [{\citenamefont {Schulz}\ \emph {et~al.}(2012{\natexlab{b}})\citenamefont {Schulz}, \citenamefont {Ritz}, \citenamefont {Bauer}, \citenamefont {Halder}, \citenamefont {Wagner}, \citenamefont {Franz}, \citenamefont {Pfleiderer}, \citenamefont {Everschor}, \citenamefont {Garst},\ and\ \citenamefont {Rosch}}]{Schulz2012-fw}%
  \BibitemOpen
  \bibfield  {author} {\bibinfo {author} {\bibfnamefont {T.}~\bibnamefont {Schulz}}, \bibinfo {author} {\bibfnamefont {R.}~\bibnamefont {Ritz}}, \bibinfo {author} {\bibfnamefont {A.}~\bibnamefont {Bauer}}, \bibinfo {author} {\bibfnamefont {M.}~\bibnamefont {Halder}}, \bibinfo {author} {\bibfnamefont {M.}~\bibnamefont {Wagner}}, \bibinfo {author} {\bibfnamefont {C.}~\bibnamefont {Franz}}, \bibinfo {author} {\bibfnamefont {C.}~\bibnamefont {Pfleiderer}}, \bibinfo {author} {\bibfnamefont {K.}~\bibnamefont {Everschor}}, \bibinfo {author} {\bibfnamefont {M.}~\bibnamefont {Garst}},\ and\ \bibinfo {author} {\bibfnamefont {A.}~\bibnamefont {Rosch}},\ }\href@noop {} {\bibfield  {journal} {\bibinfo  {journal} {Nat. Phys.}\ }\textbf {\bibinfo {volume} {8}},\ \bibinfo {pages} {301} (\bibinfo {year} {2012}{\natexlab{b}})}\BibitemShut {NoStop}%
\bibitem [{\citenamefont {Nakabayashi}\ and\ \citenamefont {Tatara}(2014)}]{Nakabayashi_2014}%
  \BibitemOpen
  \bibfield  {author} {\bibinfo {author} {\bibfnamefont {N.}~\bibnamefont {Nakabayashi}}\ and\ \bibinfo {author} {\bibfnamefont {G.}~\bibnamefont {Tatara}},\ }\href {https://doi.org/10.1088/1367-2630/16/1/015016} {\bibfield  {journal} {\bibinfo  {journal} {New J. Phys.}\ }\textbf {\bibinfo {volume} {16}},\ \bibinfo {pages} {015016} (\bibinfo {year} {2014})}\BibitemShut {NoStop}%
\bibitem [{\citenamefont {Yang}\ and\ \citenamefont {Mills}(1954)}]{PhysRev.96.191}%
  \BibitemOpen
  \bibfield  {author} {\bibinfo {author} {\bibfnamefont {C.~N.}\ \bibnamefont {Yang}}\ and\ \bibinfo {author} {\bibfnamefont {R.~L.}\ \bibnamefont {Mills}},\ }\href {https://doi.org/10.1103/PhysRev.96.191} {\bibfield  {journal} {\bibinfo  {journal} {Phys. Rev.}\ }\textbf {\bibinfo {volume} {96}},\ \bibinfo {pages} {191} (\bibinfo {year} {1954})}\BibitemShut {NoStop}%
\bibitem [{\citenamefont {Kjaergaard}\ \emph {et~al.}(2012)\citenamefont {Kjaergaard}, \citenamefont {W\"olms},\ and\ \citenamefont {Flensberg}}]{PhysRevB.85.020503}%
  \BibitemOpen
  \bibfield  {author} {\bibinfo {author} {\bibfnamefont {M.}~\bibnamefont {Kjaergaard}}, \bibinfo {author} {\bibfnamefont {K.}~\bibnamefont {W\"olms}},\ and\ \bibinfo {author} {\bibfnamefont {K.}~\bibnamefont {Flensberg}},\ }\href {https://doi.org/10.1103/PhysRevB.85.020503} {\bibfield  {journal} {\bibinfo  {journal} {Phys. Rev. B}\ }\textbf {\bibinfo {volume} {85}},\ \bibinfo {pages} {020503} (\bibinfo {year} {2012})}\BibitemShut {NoStop}%
\bibitem [{\citenamefont {Braunecker}\ \emph {et~al.}(2010)\citenamefont {Braunecker}, \citenamefont {Japaridze}, \citenamefont {Klinovaja},\ and\ \citenamefont {Loss}}]{PhysRevB.82.045127}%
  \BibitemOpen
  \bibfield  {author} {\bibinfo {author} {\bibfnamefont {B.}~\bibnamefont {Braunecker}}, \bibinfo {author} {\bibfnamefont {G.~I.}\ \bibnamefont {Japaridze}}, \bibinfo {author} {\bibfnamefont {J.}~\bibnamefont {Klinovaja}},\ and\ \bibinfo {author} {\bibfnamefont {D.}~\bibnamefont {Loss}},\ }\href {https://doi.org/10.1103/PhysRevB.82.045127} {\bibfield  {journal} {\bibinfo  {journal} {Phys. Rev. B}\ }\textbf {\bibinfo {volume} {82}},\ \bibinfo {pages} {045127} (\bibinfo {year} {2010})}\BibitemShut {NoStop}%
\bibitem [{\citenamefont {Groth}\ \emph {et~al.}(2014)\citenamefont {Groth}, \citenamefont {Wimmer}, \citenamefont {Akhmerov},\ and\ \citenamefont {Waintal}}]{Groth_2014}%
  \BibitemOpen
  \bibfield  {author} {\bibinfo {author} {\bibfnamefont {C.~W.}\ \bibnamefont {Groth}}, \bibinfo {author} {\bibfnamefont {M.}~\bibnamefont {Wimmer}}, \bibinfo {author} {\bibfnamefont {A.~R.}\ \bibnamefont {Akhmerov}},\ and\ \bibinfo {author} {\bibfnamefont {X.}~\bibnamefont {Waintal}},\ }\href {https://doi.org/10.1088/1367-2630/16/6/063065} {\bibfield  {journal} {\bibinfo  {journal} {New J. Phys.}\ }\textbf {\bibinfo {volume} {16}},\ \bibinfo {pages} {063065} (\bibinfo {year} {2014})}\BibitemShut {NoStop}%
\bibitem [{\citenamefont {Nikoli\ifmmode~\acute{c}\else \'{c}\fi{}}\ \emph {et~al.}(2005)\citenamefont {Nikoli\ifmmode~\acute{c}\else \'{c}\fi{}}, \citenamefont {Souma}, \citenamefont {Z\^arbo},\ and\ \citenamefont {Sinova}}]{PhysRevLett.95.046601}%
  \BibitemOpen
  \bibfield  {author} {\bibinfo {author} {\bibfnamefont {B.~K.}\ \bibnamefont {Nikoli\ifmmode~\acute{c}\else \'{c}\fi{}}}, \bibinfo {author} {\bibfnamefont {S.}~\bibnamefont {Souma}}, \bibinfo {author} {\bibfnamefont {L.~P.}\ \bibnamefont {Z\^arbo}},\ and\ \bibinfo {author} {\bibfnamefont {J.}~\bibnamefont {Sinova}},\ }\href {https://doi.org/10.1103/PhysRevLett.95.046601} {\bibfield  {journal} {\bibinfo  {journal} {Phys. Rev. Lett.}\ }\textbf {\bibinfo {volume} {95}},\ \bibinfo {pages} {046601} (\bibinfo {year} {2005})}\BibitemShut {NoStop}%
\bibitem [{\citenamefont {Nagaosa}\ \emph {et~al.}(2010)\citenamefont {Nagaosa}, \citenamefont {Sinova}, \citenamefont {Onoda}, \citenamefont {MacDonald},\ and\ \citenamefont {Ong}}]{RevModPhys.82.1539}%
  \BibitemOpen
  \bibfield  {author} {\bibinfo {author} {\bibfnamefont {N.}~\bibnamefont {Nagaosa}}, \bibinfo {author} {\bibfnamefont {J.}~\bibnamefont {Sinova}}, \bibinfo {author} {\bibfnamefont {S.}~\bibnamefont {Onoda}}, \bibinfo {author} {\bibfnamefont {A.~H.}\ \bibnamefont {MacDonald}},\ and\ \bibinfo {author} {\bibfnamefont {N.~P.}\ \bibnamefont {Ong}},\ }\href {https://doi.org/10.1103/RevModPhys.82.1539} {\bibfield  {journal} {\bibinfo  {journal} {Rev. Mod. Phys.}\ }\textbf {\bibinfo {volume} {82}},\ \bibinfo {pages} {1539} (\bibinfo {year} {2010})}\BibitemShut {NoStop}%
\bibitem [{\citenamefont {An}\ \emph {et~al.}(2012)\citenamefont {An}, \citenamefont {Liu}, \citenamefont {Lin},\ and\ \citenamefont {Liu}}]{An2012-da}%
  \BibitemOpen
  \bibfield  {author} {\bibinfo {author} {\bibfnamefont {Z.}~\bibnamefont {An}}, \bibinfo {author} {\bibfnamefont {F.~Q.}\ \bibnamefont {Liu}}, \bibinfo {author} {\bibfnamefont {Y.}~\bibnamefont {Lin}},\ and\ \bibinfo {author} {\bibfnamefont {C.}~\bibnamefont {Liu}},\ }\href@noop {} {\bibfield  {journal} {\bibinfo  {journal} {Sci. Rep.}\ }\textbf {\bibinfo {volume} {2}},\ \bibinfo {pages} {388} (\bibinfo {year} {2012})}\BibitemShut {NoStop}%
\bibitem [{\citenamefont {Manchon}\ \emph {et~al.}(2015)\citenamefont {Manchon}, \citenamefont {Koo}, \citenamefont {Nitta}, \citenamefont {Frolov},\ and\ \citenamefont {Duine}}]{Manchon2015-ds}%
  \BibitemOpen
  \bibfield  {author} {\bibinfo {author} {\bibfnamefont {A.}~\bibnamefont {Manchon}}, \bibinfo {author} {\bibfnamefont {H.~C.}\ \bibnamefont {Koo}}, \bibinfo {author} {\bibfnamefont {J.}~\bibnamefont {Nitta}}, \bibinfo {author} {\bibfnamefont {S.~M.}\ \bibnamefont {Frolov}},\ and\ \bibinfo {author} {\bibfnamefont {R.~A.}\ \bibnamefont {Duine}},\ }\href@noop {} {\bibfield  {journal} {\bibinfo  {journal} {Nat. Mat.}\ }\textbf {\bibinfo {volume} {14}},\ \bibinfo {pages} {871} (\bibinfo {year} {2015})}\BibitemShut {NoStop}%
\bibitem [{\citenamefont {Birch}\ \emph {et~al.}(2020)\citenamefont {Birch}, \citenamefont {Cort{\'e}s-Ortu{\~n}o}, \citenamefont {Turnbull}, \citenamefont {Wilson}, \citenamefont {Gro{\ss}}, \citenamefont {Tr{\"a}ger}, \citenamefont {Laurenson}, \citenamefont {Bukin}, \citenamefont {Moody}, \citenamefont {Weigand}, \citenamefont {Sch{\"u}tz}, \citenamefont {Popescu}, \citenamefont {Fan}, \citenamefont {Steadman}, \citenamefont {Verezhak}, \citenamefont {Balakrishnan}, \citenamefont {Loudon}, \citenamefont {Twitchett-Harrison}, \citenamefont {Hovorka}, \citenamefont {Fangohr}, \citenamefont {Ogrin}, \citenamefont {Gr{\"a}fe},\ and\ \citenamefont {Hatton}}]{Birch2020-rj}%
  \BibitemOpen
  \bibfield  {author} {\bibinfo {author} {\bibfnamefont {M.~T.}\ \bibnamefont {Birch}}, \bibinfo {author} {\bibfnamefont {D.}~\bibnamefont {Cort{\'e}s-Ortu{\~n}o}}, \bibinfo {author} {\bibfnamefont {L.~A.}\ \bibnamefont {Turnbull}}, \bibinfo {author} {\bibfnamefont {M.~N.}\ \bibnamefont {Wilson}}, \bibinfo {author} {\bibfnamefont {F.}~\bibnamefont {Gro{\ss}}}, \bibinfo {author} {\bibfnamefont {N.}~\bibnamefont {Tr{\"a}ger}}, \bibinfo {author} {\bibfnamefont {A.}~\bibnamefont {Laurenson}}, \bibinfo {author} {\bibfnamefont {N.}~\bibnamefont {Bukin}}, \bibinfo {author} {\bibfnamefont {S.~H.}\ \bibnamefont {Moody}}, \bibinfo {author} {\bibfnamefont {M.}~\bibnamefont {Weigand}}, \bibinfo {author} {\bibfnamefont {G.}~\bibnamefont {Sch{\"u}tz}}, \bibinfo {author} {\bibfnamefont {H.}~\bibnamefont {Popescu}}, \bibinfo {author} {\bibfnamefont {R.}~\bibnamefont {Fan}}, \bibinfo {author} {\bibfnamefont {P.}~\bibnamefont {Steadman}}, \bibinfo {author} {\bibfnamefont {J.~A.~T.}\ \bibnamefont {Verezhak}}, \bibinfo {author}
  {\bibfnamefont {G.}~\bibnamefont {Balakrishnan}}, \bibinfo {author} {\bibfnamefont {J.~C.}\ \bibnamefont {Loudon}}, \bibinfo {author} {\bibfnamefont {A.~C.}\ \bibnamefont {Twitchett-Harrison}}, \bibinfo {author} {\bibfnamefont {O.}~\bibnamefont {Hovorka}}, \bibinfo {author} {\bibfnamefont {H.}~\bibnamefont {Fangohr}}, \bibinfo {author} {\bibfnamefont {F.~Y.}\ \bibnamefont {Ogrin}}, \bibinfo {author} {\bibfnamefont {J.}~\bibnamefont {Gr{\"a}fe}},\ and\ \bibinfo {author} {\bibfnamefont {P.~D.}\ \bibnamefont {Hatton}},\ }\href@noop {} {\bibfield  {journal} {\bibinfo  {journal} {Nat. Com.}\ }\textbf {\bibinfo {volume} {11}},\ \bibinfo {pages} {1726} (\bibinfo {year} {2020})}\BibitemShut {NoStop}%
\bibitem [{\citenamefont {Hus}\ \emph {et~al.}(2017)\citenamefont {Hus}, \citenamefont {Zhang}, \citenamefont {Nguyen}, \citenamefont {Ko}, \citenamefont {Baddorf}, \citenamefont {Chen},\ and\ \citenamefont {Li}}]{PhysRevLett.119.137202}%
  \BibitemOpen
  \bibfield  {author} {\bibinfo {author} {\bibfnamefont {S.~M.}\ \bibnamefont {Hus}}, \bibinfo {author} {\bibfnamefont {X.-G.}\ \bibnamefont {Zhang}}, \bibinfo {author} {\bibfnamefont {G.~D.}\ \bibnamefont {Nguyen}}, \bibinfo {author} {\bibfnamefont {W.}~\bibnamefont {Ko}}, \bibinfo {author} {\bibfnamefont {A.~P.}\ \bibnamefont {Baddorf}}, \bibinfo {author} {\bibfnamefont {Y.~P.}\ \bibnamefont {Chen}},\ and\ \bibinfo {author} {\bibfnamefont {A.-P.}\ \bibnamefont {Li}},\ }\href {https://doi.org/10.1103/PhysRevLett.119.137202} {\bibfield  {journal} {\bibinfo  {journal} {Phys. Rev. Lett.}\ }\textbf {\bibinfo {volume} {119}},\ \bibinfo {pages} {137202} (\bibinfo {year} {2017})}\BibitemShut {NoStop}%
\bibitem [{\citenamefont {B\"uttiker}(1986)}]{PhysRevLett.57.1761}%
  \BibitemOpen
  \bibfield  {author} {\bibinfo {author} {\bibfnamefont {M.}~\bibnamefont {B\"uttiker}},\ }\href {https://doi.org/10.1103/PhysRevLett.57.1761} {\bibfield  {journal} {\bibinfo  {journal} {Phys. Rev. Lett.}\ }\textbf {\bibinfo {volume} {57}},\ \bibinfo {pages} {1761} (\bibinfo {year} {1986})}\BibitemShut {NoStop}%
\bibitem [{\citenamefont {Tretiakov}\ and\ \citenamefont {Tchernyshyov}(2007)}]{PhysRevB.75.012408}%
  \BibitemOpen
  \bibfield  {author} {\bibinfo {author} {\bibfnamefont {O.~A.}\ \bibnamefont {Tretiakov}}\ and\ \bibinfo {author} {\bibfnamefont {O.}~\bibnamefont {Tchernyshyov}},\ }\href {https://doi.org/10.1103/PhysRevB.75.012408} {\bibfield  {journal} {\bibinfo  {journal} {Phys. Rev. B}\ }\textbf {\bibinfo {volume} {75}},\ \bibinfo {pages} {012408} (\bibinfo {year} {2007})}\BibitemShut {NoStop}%
\end{thebibliography}%

\end{document}